\documentclass[sigconf]{acmart}

\usepackage{csquotes}

\usepackage{subcaption}
\usepackage{graphicx} 

\definecolor{Orange}{rgb}{1,0.5,0} 
\newcommand{\revise}[1]{\textrm{\textrm{\textcolor{black}{#1}}}}

\newcommand{\zm}[1]{\textcolor{black}{#1}}

\newcommand{\yqw}[1]{\textcolor{black}{#1}}

\newcommand{\nb}[1]{\textcolor{black}{#1}}

\AtBeginDocument{%
  \providecommand\BibTeX{{%
    \normalfont B\kern-0.5em{\scshape i\kern-0.25em b}\kern-0.8em\TeX}}}

\copyrightyear{2026}
\acmYear{2026}
\setcopyright{cc}
\setcctype{by}
\acmConference[CHI '26]{Proceedings of the 2026 CHI Conference on Human Factors in Computing Systems}{April 13--17, 2026}{Barcelona, Spain}
\acmBooktitle{Proceedings of the 2026 CHI Conference on Human Factors in Computing Systems (CHI '26), April 13--17, 2026, Barcelona, Spain}
\acmPrice{}
\acmDOI{10.1145/3772318.3790678}
\acmISBN{979-8-4007-2278-3/2026/04}

\begin{document}

\title[A Meat-Summer Night's Dream]{A Meat-Summer Night's Dream: A Tangible Design Fiction Exploration of Eating Biohybrid Flying Robots}


\author{Ziming Wang}
\authornote{The author was a visiting researcher at Stanford University while part of this work was conducted.}
\affiliation{%
  \institution{Chalmers University of Technology}
  \city{Gothenburg}
  \country{Sweden}}
\affiliation{%
  \institution{University of Luxembourg}
  \city{Esch-sur-Alzette}
  \country{Luxembourg}}
\email{ziming@chalmers.se}
\orcid{0000-0003-0564-8757}

\author{Yiqian Wu}
\affiliation{%
  \institution{Chalmers University of Technology}
  \city{Gothenburg}
  \country{Sweden}}
\email{yiqianwuu@gmail.com}
\orcid{0009-0007-9105-0108}

\author{Qingxiao Zheng}
\affiliation{%
  \institution{University at Buffalo}
  \city{Buffalo}
  \state{New York}
  \country{USA}}
\email{qingxiao@buffalo.edu}
\orcid{0000-0003-0368-0032}

\author{Shihan Zhang}
\affiliation{%
  \institution{alter+ (Alter Plus)}
  \city{San Francisco}
  \state{California}
  \country{USA}}
\email{shihanzhang.design@gmail.com}
\orcid{0009-0003-0816-2813}

\author{Ned Barker}
\affiliation{%
  \institution{King's College London}
  \city{London}
  \country{United Kingdom}}
\email{edmund.barker@kcl.ac.uk}
\orcid{0000-0001-6969-4547}

\author{Morten Fjeld}
\affiliation{%
  \institution{Chalmers University of Technology}
  \city{Gothenburg}
  \country{Sweden}}
  \affiliation{%
  \institution{University of Bergen}
  \city{Bergen}
  \country{Norway}}
\email{fjeld@chalmers.se}
\orcid{0000-0002-9562-5147}

\renewcommand{\shortauthors}{Ziming Wang et al.}

\begin{abstract}

\textit{What if future dining involved eating robots?} We explore this question through a playful and poetic experiential dinner theater: a tangible design fiction staged as a 2052 Paris restaurant where diners consume a biohybrid flying robot in place of the banned delicacy of ortolan bunting. Moving beyond textual or visual speculation, our “dinner-in-the-drama” combined performance, ritual, and multisensory immersion to provoke reflection on sustainability, ethics, and cultural identity. Six participants from creative industries engaged as diners and role-players, responding with curiosity, discomfort, and philosophical debate. They imagined biohybrids as both plausible and unsettling—raising questions of sentience, symbolism, and technology adoption that extend beyond conventional sustainability framings of synthetic meat. Our contributions to HCI are threefold: (i) a speculative artifact that stages robots as food, (ii) empirical insights into how people negotiate cultural and ethical boundaries in post-natural eating, and (iii) a methodological advance in embodied, multisensory design fiction.
\end{abstract}



\begin{teaserfigure}
  \centering
  \includegraphics[width=0.939\textwidth]{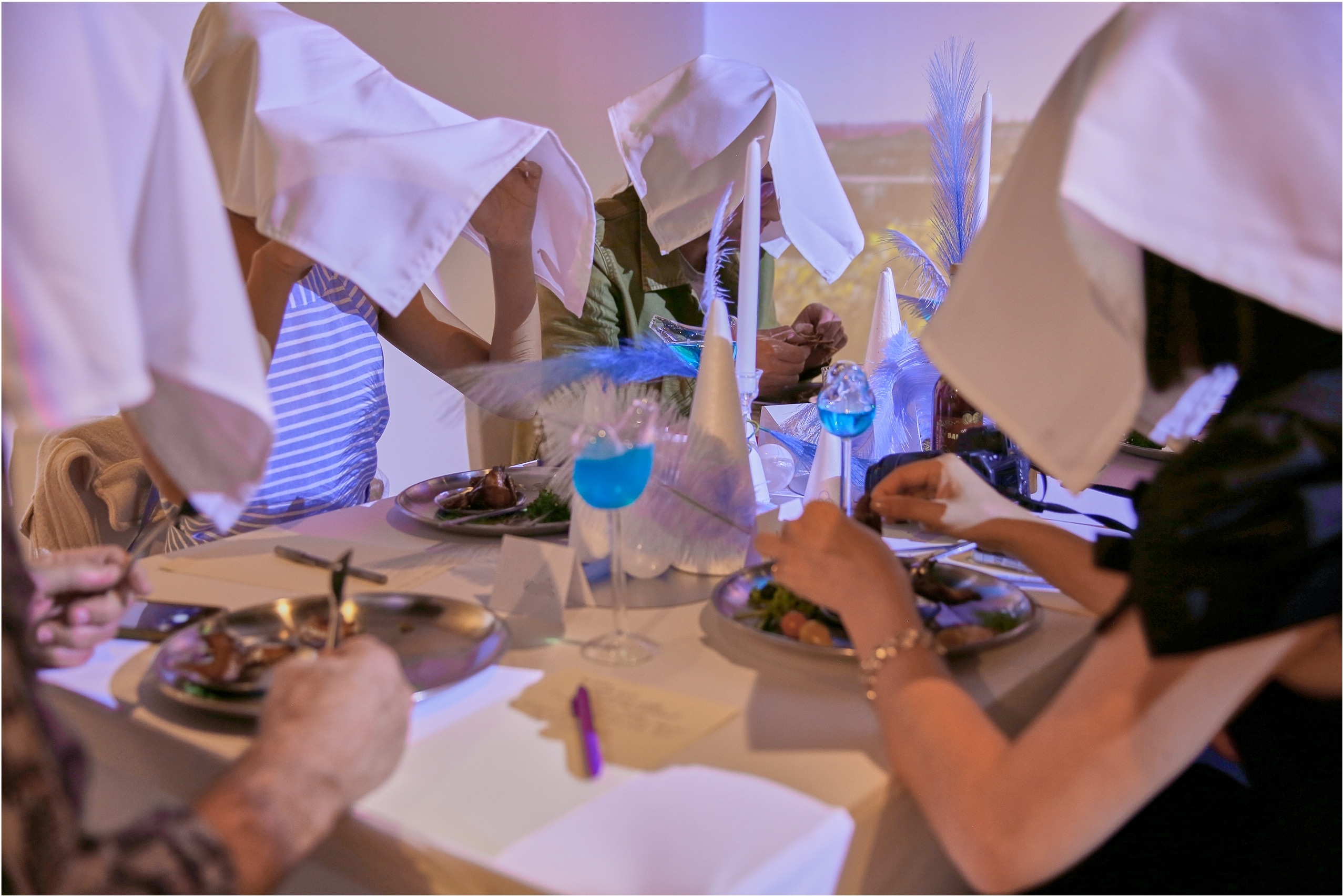}
  \caption{Participants were eating ``biohybrid flying robots'' in the traditional French manner.}
  \Description{A group of people eating at a table with napkins covering their heads.}
  \label{fig:teaser}
\end{teaserfigure}

\maketitle

\section{Introduction}\label{intro}

\begin{displayquote}
\textit{``Meat! We are going to eat some meat; and what meat!''}    \\ \hspace*{\fill}— Jules Verne,\\ \hspace*{\fill} French writer,\\ \hspace*{\fill} ``father of science fiction''. \\

\textit{``The brain is merely a meat machine.''}    \\ \hspace*{\fill}— Marvin Minsky,\\ \hspace*{\fill}American cognitive and computer scientist,\\ \hspace*{\fill} AI pioneer.\\

\end{displayquote}

\noindent Since time immemorial, humans have eaten animals. This act---mundane yet deeply cultural---has shaped our ethics, our rituals, and our relationship to the living world. Today, however, climate change, industrial farming, and shifting cultural values press us to imagine new food futures. Increasingly, researchers are exploring radically alternative possibilities. For example, the ROBOFOOD project \cite{RoboFoodResearch} exemplifies a broader movement where robotics and food science converge around the idea of edible robotics to reimagine what counts as food. \zm{These shifts create new ethical and cultural tensions that invite HCI to interrogate how technological interventions might reshape long-standing food rituals and the meaning of “eating” itself.} We ask: \textit{what if we could eat biohybrid robots---robots that are (partly) made from cultured tissues---instead of cows, chickens, or fish?} In this paper, we take this speculative question constructively, not to propose a technological roadmap, but to use it as a lens: a way to probe how emerging technologies might reshape human ethics, cultural identity, and the boundaries of human--robot interaction. 

Although eating animals has been a constant throughout evolutionary epochs, successive sociotechnical revolutions have transformed this relationship. Technologies have taken us from hunting wild animals to domestication, selective breeding \cite{humanihis}, and eventually mass production. More recently, an emergent technology of cultivated or lab-grown meat has reached commercial approval. At present, Singapore and the USA are the only two countries that have legalized the sale of lab-grown meat for human consumption, with others likely to follow. For instance, the UK has approved lab-grown meat for pet food \cite{ft_lab_grown_meat_2024}, and the Netherlands permits it for tasting purposes \cite{iamsterdam_cultivated_meat_2023}. These developments mark the beginning of a food revolution \cite{FDA}, one that seeks to address the environmental and ethical costs of industrial meat production. 

Yet the state of the art remains at an early stage, with technical, socio-political, and regulatory challenges ahead \cite{stephens_disilvio_dunsford_ellis_glencross_sexton_2018}. A key difficulty lies in replicating the form, texture, and appearance of conventional animal meat \cite{9592237}, given the complexity of biological structures \cite{griffith_naughton_2002}. This leads us to ask: \textit{Why not consider creating a biohybrid robot that leverages natural processes to grow authentic animal components, while being managed and controlled by computers?} Here, Minsky’s metaphor of the brain as a “meat machine” offers a useful conceptual pivot. By framing the brain---the seat of consciousness---as mechanistically reproducible, Minsky underscored a materialist view that, in principle, no natural phenomenon is beyond technological replication. Extending this reasoning, one might imagine not only replicating neural systems but also reconstituting organic tissues, organs, and even entire animal forms within artificial substrates. This conceptual foundation supports the possibility of assembling and eating animal--robot hybrids \cite{robotmeat}: entities that merge biological growth with computational control. From this perspective, such biohybrid robots are not merely a technical workaround for the shortcomings of current lab-grown meat. They represent a deeper continuity with cultural traditions of animals-as-food, while simultaneously reframing the act of eating within the context of human--robot interaction.

To bring this vision to life, we designed an experiential dinner theater titled \textit{A Meat-Summer Night’s Dream}, set in a fictional 2052 Paris restaurant. Six participants from creative and cultural fields engaged as diners and semi-scripted role-players \zm{as they encountered a series of tangible speculative artifacts—including a feathered drone, a translucent anatomical model, and an edible surrogate bird (see \autoref{fig:props}). Unlike conventional design fictions that rely on texts or images, our approach foregrounds multisensory experience—taste, smell, touch, sound, and visual occlusion—to elicit situated ethical reflection.} \zm{We additionally foreground the cultural stakes of eating by reimagining} the controversial French delicacy of Ortolan Bunting, substituting the bird with synthetic meat grown as part of biohybrid flying robots. \zm{This choice is not celebratory; rather, it surfaces questions of ritual, taboo, symbolism, and cultural identity that are often under-examined in HCI discussions of food futures.} To guide our inquiry, we posed two research questions: \textbf{RQ1}: How do people contemplate the idea of eating biohybrid robots when \zm{placed within a culturally and sensorially rich speculative future}? \textbf{RQ2}: How does a tangible multisensory dining experience provoke \zm{ethical, cultural, and affective reflection about post-natural} food futures?

This paper contributes to HCI \zm{in three ways}: (i) \zm{a situated, embodied} speculative artifact \zm{that employs sensory, material, and ritual elements to explore possible interactions with edible biohybrid robots}; (ii) empirical finding \zm{that illuminate how participants reason about animacy, ethical ambiguity, and cultural meaning within a multisensory speculative encounter}; and (iii) a methodological \zm{refinement of design fiction, demonstrating how} a tangible, embodied format can foster critical, experiential engagement with speculative technologies.

Finally, we clarify the scope of the work---this study does not aim for generalizability or predictive claims. Rather, its value lies in documenting how speculative dining can elicit situated ethical and cultural reasoning in an intimate, embodied context. We recognize that this paper’s provocative nature has sparked and may continue to spark diverse reactions, and we deeply appreciate this range of perspectives. We argue that such plurality is itself evidence of the value of speculative and experimental work in HCI, namely its capacity to open questions, stimulate discussion, and broaden our imagination of what interaction can be.

\section{Background and Related Work}


\subsection{New Frontiers for Food and Technology}\label{hfi}

\subsubsection{\textbf{Human--Food Interaction in HCI}}
Food is not only essential for survival, it is often a deeply social and cultural experience that can be enjoyable, unusual or repulsive \cite{khot2019human}. Technology has crucially supported and enriched food-related practices \cite{khot2019human}. \citet{grime08} opened a discussion on the role of food-tech in HCI by introducing speculative design concepts for edible interfaces. As the intersection of food and technology has been explored, there has been a growing trend in exploring Human--Food Interaction (HFI) to create, transform, and elevate food-related experiences with technology \cite{karunanayaka, spence2014}. This research has taken two directions: “around food” and “with food” \cite{gay20}. While design “around food” focuses more on the social experience of consuming food, design “with food” is oriented to crafting edible user experiences. A number of recent  projects have involved the growth, preparation, and consumption of new types of foods \cite{wangw17, zoran18, kan} and has delved into speculative food-tech applications, imagining how technological innovations could transform the sensory experience of eating \cite{CHI24FoodLeaf, CHI24foodFupop}. In another technological innovation, Chen et al. examined ways to foster people's affective emotion toward fermentative microbes \cite{Chen2021}, proposing Nukabot—a conversational artifact supporting interaction between microbes and humans through maintenance, affection, and obligation. Relatedly, edible robotic systems have been envisioned as food that not only serves nutritional purposes but also embodies sensing, actuation, and computational functions, enabling applications from targeted drug delivery to ecological intervention \cite{floreano2024edible}. 

However, the space of design “with food” has been less explored compared to design “around food” due to its challenging nature of going beyond traditional screen-based products \cite{obr16} and underutilization of taste- and smell-based interfaces within the field of HCI \cite{obrist, obrist14, rana11}. Furthermore, existing studies have emphasized social and cultural aspects of technology in dining \cite{grime08, hup12}, rather than aiming to improve the individual eating experience \cite{altar19}. \yqw{To bridge these perspectives, Deng et al. proposed nine HFI building blocks \cite{deng2023}, along with tangible “Dancing Delicacies” artifacts \cite{deng2023dancing}, contributing new methods and material vocabularies for HFI. Their concept of ``computational food'' \cite{deng2024} emphasized the choreographic potential of interactive foods and systematic study of “complex interactions among creators, computational food, and consumers.” Other projects oriented toward fundamental dining experiences include Logic Bonbon \cite{deng22}, Chewing Jockey \cite{tanaka2011}, and LoLLio \cite{murer13}.}

Despite growing activity in advancing both HFI directions, there has been limited exploration of food sources and food production technology. Although \citet{robotmeat} proposed a speculative idea around consuming animal-robot hybrids, empirical HCI studies that test such concepts are scarce in this area of research and development. \yqw{This study addresses this gap by introducing \textit{tangible speculative food artifacts} to explore new imaginaries of sourcing and consuming food, and to provoke reflection on the ethical and affective entanglements involved in post-natural eating.}

\subsubsection{\textbf{Synthetic Food: Cultivated Meat and Its Limitations}}\label{lgm}
Frontier developments in bioscience and bioengineering, such as cellular agriculture and tissue engineering, have enabled the in-vitro cultivation of animal cells to produce edible meat products. Cultured or lab-grown meat is often promoted as a “slaughter-free”, environmentally sustainable alternative to conventional animal agriculture \cite{9592237}. Its purported benefits include reduced ecological impact, elimination of animal suffering, and improved control over foodborne diseases.  

Despite this promise, cultivated meat faces significant scientific and socio-political barriers. Key technical limitations include replicating the complex, hierarchical structures of natural tissues, as integrating microscale functional units into macroscale edible forms remains an ongoing challenge \cite{griffith_naughton_2002}. Furthermore, regulatory, cultural, and consumer acceptance issues persist \cite{stephens_disilvio_dunsford_ellis_glencross_sexton_2018}.

\yqw{These limitations open space for alternative visions. We suggest that integrating robotics into tissue-based food systems—i.e., constructing animal-robot hybrids (see~\autoref{concept}), offers speculative, yet provocative pathways to address structural, sensory, and ontological challenges in synthetic food futures. In parallel, advances in edible robotics, such as edible drones \cite{kwak2022edible} and biodegradable aquatic robots \cite{zhang2025edible}, illustrate how food and robotic systems might merge to provide both functional performance and nutritional value in extreme or resource-limited environments, suggesting provocative future scenarios for post-natural eating.}

\subsection{The Idea of Living Machines, Biohybrid Robots, and the Animal-Robot Analogy}
Over the last decade, the term Living Machines has been employed to signal a growing convergence between biology and technology (see \cite{10.1117/12.2046305, 10.1093/oso/9780199674923.001.0001}). In both scientific and artistic domains, researchers have developed machines inspired by biological forms and processes \cite{naturalsoundandproxemics, PCI, flapperdrone, inaflap, chi24drone, ZWang_PhD}, and increasingly, machines built from living tissue itself \cite{Shoji2018, KAWAI2024102066, KINJO2024948}. For instance, Takeuchi and colleagues created a biohybrid bipedal robot powered by skeletal muscle tissue \cite{KINJO2024948} and a humanoid robot covered with living skin capable of smiling \cite{KAWAI2024102066}. Other works employ fungi as computational substrates for robotic control \cite{cornellmush}, blurring the boundaries between machine and organism. At the same time, ethical challenges and responsibilities in biohybrid robotics research have been raised \cite{biohybrid_ethics, kirksey2021living}, as these efforts increasingly collapse distinctions between the biological and the mechanical. These efforts are part of a broader movement to develop biohybrid systems, where biological and synthetic components form new composite entities \cite{Jung11, Webster-Wood_2023}. \yqw{Within this movement, edible robotic concepts \cite{floreano2024edible} also extend the scope of “living” or “nutritive” machines, challenging the division between machine, organism, and food.}

As a significant cultural metaphor, Living Machines build upon an idea that humans will continue to radically control and re-engineer the living world, including animals, to bend to our will \cite{vaage}. Some advocate such systems as ecologically viable alternatives to resource-intensive AI and robotics \cite{10.1093/oso/9780199674923.003.0065}. Relatedly, animal-robot analogies have been explored in Human--Robot Interaction (HRI), suggesting that robots may be treated “like animals” to supplement social, cognitive, and emotional roles in human life \cite{coeck, juyweng, nehanivdautenhahn2007, darling2021new}. Previous research also suggests that people perceive machines as possessing consciousness to some extent \cite{scott2023do, Zhang23, Xu18}, which further complicates how living qualities are projected onto artificial systems. Complementing these techno-scientific efforts, artistic works such as \textit{Lapillus Bug}\cite{lapillusbug} similarly explore the animacy of inorganic materials, blurring boundaries between life and non-life through speculative, sensory-driven design.
 
Against this backdrop of humans' propensity to exploit animals, two main topics emerge: harvesting animal products for food, and demanding services from animals. While many investigations have explored animals and robots through the lens of service provision, the possibility of cultivating animal products through robotic means remains largely unexamined in HRI and robotics research. \yqw{This absence may stem from earlier technical and conceptual constraints—when animals were strictly biological and robots merely mechanical. Yet the emergence of Living Machines collapses these boundaries, expanding the sociotechnical horizon to include speculative systems where food, life, and machinery converge. It also prompts a deeper question: \emph{What implications might such hybrids have for other human--robot and human--animal relations?}}

\subsection{Speculative Design and Design Fictions in HCI}
Speculative design is a cover term for various forms of design fiction and critical design \cite{cameraCars2023}, we consider they are interchangeable. \yqw{Design fiction has become an increasingly influential approach in HCI—particularly within research through design \cite{resthrdes, designfictionCHI20EA}. As \citet{dunne2024speculative} described, speculative design serves as “a catalyst for social dreaming”, enabling the creation of sociotechnical imaginaries \cite{soctechimag} that challenge existing paradigms. Rather than solving problems, design fiction invites reflection and provokes debate \cite{Forlano02102014, galloway2018speculative, criticaldesign}, often using wit or satire to spark critical engagement \cite{Malpass01112013}.}

\yqw{Recent work has demonstrated speculative design's potential across a variety of domains. In robotics and AI, design fiction has been used to anticipate ethical challenges \cite{ringfort2023} and foreground public accountability in surveillance systems \cite{cameraCars2023}. Within participatory and justice-oriented frameworks, speculative methods have empowered marginalized communities to envision alternative futures \cite{radicalafrifu, incdivspedesCHIEA, blafutCHI22}. In food-tech, speculative design has supported community engagement in sustainable food futures \cite{participatefood}, while others have explored ecological imaginaries and sustainable design through fictional scenarios \cite{susgreendesfic}. Futures studies have further expanded the temporal scope of HCI research, promoting long-term thinking and creative radicalism in design \cite{MankoffCHI13, desfutChineseCHI}.}

\yqw{While inspiring, many of these projects rely primarily on textual, visual, or virtual formats for speculation (e.g., \cite{cameraCars2023, CHI22VR, CHI24AIparent}), where participants interact passively as viewers or readers. Outside HCI, works like \textit{The Anthropocene Cookbook} \cite{AnthropoceneCookbook} similarly speculate on future food systems through artistic provocations. However, they too often remain abstract or conceptually distant. Yet human culture \cite{2025aieyes}, politics, and ethics are fundamentally embodied and multisensory. To support deeper experiential engagement, we argue for the inclusion of tangible, multisensory elements in speculative design. In this work, we adopt an approach we call \textbf{tangible design fiction}, integrating material and sensory experience into speculative world-building to enable richer, more situated encounters with imagined futures.}

\subsection{Cultures of Edibility and Biohybrid Ethics}
Researchers have long argued that food ethics and cultural practices are co-constitutive: what counts as acceptable eating is shaped by collective identities, social memories, and moral orders \cite{fischler_1988, counihan_2013, douglas_1972}. Synthetic and biohybrid foods enter these domains not as neutral technologies but as culturally interpreted matter entangled with histories of purity, taboo, kinship, and ritual \cite{Rozin2008Disgust, harris_1987, harris_2001}. Studies in food anthropology show that unfamiliar foods are typically assimilated through analogy, symbolic categorization, and everyday narrative rather than technical knowledge \cite{evan_sutton_2006, paxson_2008, mintz_1986}, indicating that interpretation is grounded in accumulated cultural repertoires and affective attachments.

Affective and sensory dimensions are central to this negotiation. Previous works \cite{probyn2016eating, highmore_2008} show that disgust, curiosity, care, and desire stabilize decisions about edibility. Multispecies ethics research further argues that judgments about consuming hybrid organisms hinge on perceived agency, liveliness, and relationality \cite{donnaharaway_2008, braidotti_2013}. The introduction of semi-living tissues and biohybrid robotics complicates conventional distinctions between life and mechanism, raising questions familiar from stem-cell ethics, embryology, and synthetic biology \cite{hyun2016embryology, internationalsocietyforstemcellresearch_2021, Webster-Wood_2023}.

STS scholarship highlights that ethical reasoning around food technologies is embedded in broader political and ecological imaginaries. Analyses of food futures emphasize how visions of climate adaptation, industrial scaling, and geopolitical distribution shape acceptance of synthetic and lab-grown products \cite{alice_dal_gobbo_2023, ayakannu_vijayan_elengoe_2024, tamás_landesz_2023}. Paxson’s post-pasteurian microbiopolitics \cite{paxson_2008} shows that novel food systems produce new forms of governance, expertise, and public trust. Star and Griesemer’s concept of boundary objects \cite{star_griesemer_1989} has been applied to food technologies as hybrid entities negotiated across scientific, culinary, and moral communities \cite{bingham_2006}, framing edibility as a provisional agreement rather than an intrinsic property.

Recent speculative and sensory design research provides methodological insight into how publics encounter ethical ambiguities before technologies are widely implemented. Immersive and anticipatory design methods reveal latent values and cultural scripts that remain obscure in abstract discussion \cite{auger_2013, michael_2012, jewitt_barker_steimle_2022}. Lupton and Willis \cite{lupton_willis_2021, lupton_2022} demonstrate that participatory, embodied encounters offer “anticipatory ethics”, allowing participants to rehearse future dilemmas using familiar cultural frameworks. Related multisensory design work shows that taste, smell, and ritual materially shape ethical reasoning, not merely represent it \cite{pink2023sensory}.

Existing scholarship makes clear that judgments of edibility emerge through cultural infrastructures, multispecies ethics, and affective reasoning, yet these dynamics are mostly examined retrospectively. What remains less understood is how publics might actively negotiate ethical ambiguities before biohybrid foods become commonplace. Our study addresses this gap by using speculative dining to elicit situated ethical reasoning through ritual, narrative, and sensory engagement, positioning culture as an active arena where meanings of biohybridity and synthetic eating are negotiated in real time.





\section{Methodology}

\subsection{Ideation and Concept}

\subsubsection{\textbf{General Speculative Concept of Eating Animal-Robot Hybrid}} \label{concept}
As a speculation on future food systems beyond lab-grown meat, we conceptualize a biohybrid robot composed of two major parts: (i) an artificial, computer-based “brain”, and (ii) a semi-biological body embedded with living tissue, grown from real animal stem cells. The aim is to imagine a provocative but plausible progression of current tissue engineering and robotics research, bypassing existing constraints in replicating the texture and structure of conventional meat products \cite{griffith_naughton_2002}. 

The artificial “brain” consists of AI-based electronics capable of regulating the body's internal biological systems, much like a rudimentary autonomic nervous system. The “body”, in turn, acts as a mobile incubator for stem-cell growth, allowing differentiated biological tissues—such as muscle and skin—to grow in vivo. These hybrids, while mobile and organism-like, would remain non-conscious by design (i.e., lacking a cerebrum), enabling their assembly, harvesting, and disassembly without invoking traditional ethical concerns about slaughter.

Importantly, this concept is not intended as a direct technological proposal, but as a design fiction to stimulate reflection on how future food technologies might intersect with ethics, sentience, and biological complexity. The speculative system allows for diverse configurations---where the species, morphology, and AI complexity of each hybrid can vary based on culinary or cultural interest (e.g., bird-like versus insect-like forms). In developing this concept, we drew on parallels with speculative robotics and biohybrids in research \cite{KINJO2024948, KAWAI2024102066}, as well as engaging with philosophical perspectives on consciousness, sentience, and embodied intelligence \cite{gupta2021embodied}. The biohybrid robot, in this context, acts as a vessel through which to explore boundaries between the animate and inanimate, edible and ethical, animal and machine.

\subsubsection{\textbf{Biohybrid Flying Robots as Synthetic Ortolan for Eating}}

To provoke discussion on the tensions between tradition, ethics, and technology, we designed a speculative scenario in which diners consume a biohybrid flying robot inspired by the controversial French delicacy—the Ortolan Bunting. This choice is not intended as an endorsement of the original practice, but rather as a culturally loaded and emotionally charged case that allows us to explore the potential of synthetic organisms to preserve, transform, or challenge endangered culinary rituals.

The Ortolan Bunting is a small songbird historically consumed in French haute cuisine, where it was captured, force-fed, and drowned in Armagnac before roasting—a process widely regarded as cruel \cite{tastingtable_ortolan_2022, eater_ortolans_2018}. The dining ritual, involving covering one's head with a napkin, was said to preserve aroma and also to hide the act from divine judgement. Its consumption was banned due to ethical and ecological concerns \cite{smithsonian_ortolans_2023}. Yet, its symbolic cultural resonance persists.

In our speculative future, we reimagine this delicacy through a biohybrid robot whose meat—cultured from ortolan stem cells—is attached to a flying drone body. This hybrid allows us to explore how advanced biotechnology might intersect with ritual, ethics, and sensory experience. The traditional napkin ritual is retained, not as endorsement, but as a way to question how cultural practices may persist, evolve, or become recontextualized when paired with future technologies (see \autoref{fig:teaser}). This design fiction serves as a provocation to examine broader (e.g., ethical, cultural, political) questions raised by prospects of alternative and radical technological food futures---providing a stage on which we could explore the study's two research questions (see \autoref{intro}). 


\begin{figure*}[h]
    \centering
    \includegraphics[width=\linewidth]{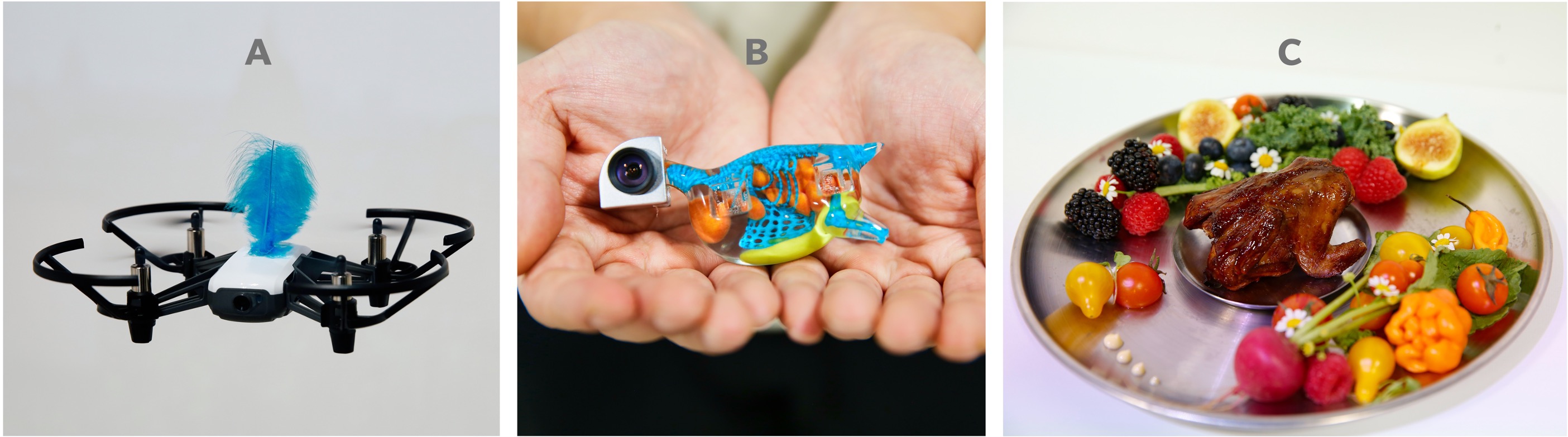} 
    \Description{From left to right: a flying robot with blue feather, a colorful resin prototype, a bird dish with fruits and vegetables on a plate.}
    \caption{Theatrical props showcasing the biohybrid flying robot in three different stages: (a) Flying drone (in motion); (b) 3D-printed prototype; (c) Roasted bird (edible).} 
    \label{fig:props}
\end{figure*}

\subsection{Dinner-in-the-Drama as Tangible Design Fiction}

Design fiction is a widely used method in speculative and critical design to explore possible futures. However, most design fiction implementations in HCI have traditionally relied on visual or textual materials, such as videos, images, or written scenarios—often limiting participant engagement to a cognitive or imaginative level. In contrast, we propose a novel methodological extension we term \textbf{tangible design fiction}, which intentionally incorporates multisensory, physically embodied elements into the fictional scenario to foster immersive, situated reflection.

Our approach is motivated by the observation that human experience, especially culinary practices, is inherently multisensory: touch, smell, sound, and taste shape how we perceive and emotionally respond to the world. By engaging these sensory modalities, tangible design fiction narrows the distance between speculative imagination and physical experience, enabling participants to feel, not just imagine, a possible future.

In this study we translated this method into what we call \textbf{``dinner-in-the-drama''}, a format that transforms traditional “dinner theater” into a participatory, performative research setting. Unlike dinner theater, where guests are passive observers of a staged narrative—\emph{dinner-in-the-drama} immerses participants as actors within the fiction. They not only consume the food but perform roles, navigate rituals, and interact with speculative artifacts, becoming part of a live, co-created scenario. This format draws inspiration from methods such as speculative enactments and live-action roleplay (LARP) in design research, but is distinct in its grounding in sensory food interaction and co-performance of situated cultural scripts.

By synthesizing speculative world-building, role play, and multisensory and cross-media design, \emph{dinner-in-the-drama} offers a novel contribution to both HCI and Human-Food Interaction. It expands the repertoire of design fiction methods and supports the embodied evaluation of speculative technologies and their sociocultural implications.



\begin{table*}[h]
\Small
\caption{Roles, personas, and suggested contributions for each participant in the drama.} 
\label{tab:p}
\begin{tabular}{p{0.16\textwidth} p{0.07\textwidth} p{0.43\textwidth} p{0.25\textwidth}}
\toprule
\textbf{Person} & \textbf{Role} & \textbf{Persona} & \textbf{Suggested Contribution} \\ \midrule
\textbf{Lead Researcher} & Chef & A knowledgeable chef, sometimes has crazy ideas. &  \\
\textbf{Lead Artist} & Host & A hospitable and rich host. &  \\
\textbf{Project Assistant} & Lensman & A quiet person, but a hardworking man. &  \\
\textbf{\revise{P1: Design Lecturer}} (30s, F) & Manager & A restaurant manager, also a researcher and designer passionate about developing creative strategies in regions affected by environmental and technological shifts. & Discuss ethical considerations of the future scenario and technology. \\
\textbf{\revise{P2: Poet}} (30s, M) & Diner & A poet with a background in literature, blending technology, food, and philosophy in his works. He plans to publish a collection exploring the intersection of these themes. & Raise philosophical questions about the scenario and technology. \\
\textbf{\revise{P3: Artist}} (30s, F) & Diner & An artist and curator in Paris, thriving on creativity and innovation at the intersection of art and technology. She is also a food enthusiast. & Guide the group on socio-cultural impacts of the future scenario. \\
\textbf{\revise{P4: Foresight Expert}} (70s, M) & Diner & A professional futurist with expertise in global technological trends. He enjoys exploring food markets and visiting innovative restaurants around the world. & Lead a discussion on economic impacts of the future technology. \\
\textbf{\revise{P5: UX Designer}} (30s, F) & Diner & A Paris-based designer who has lived in France for a decade. She is knowledgeable about French food culture. & Discuss environmental impacts of the future technology. \\
\textbf{\revise{P6: Filmmaker}} (30s, M) & Diner & A Paris-born film director passionate about cinema and culinary traditions. His work is deeply influenced by the cultural essence of Paris. & Facilitate conversation on political impacts of the technology. \\ 
\bottomrule
\end{tabular}
\end{table*}

\subsection{Drama Design and Prototyping}

The immersive environment of the dinner-in-the-drama was carefully curated to serve both narrative and methodological purposes, offering a multisensory anchor for the speculative scenario. Our goal was to embed participants in a future where cultural, ethical, and sensory boundaries around food are reimagined. Accordingly, props, atmosphere, and culinary design worked in tandem to construct a tangible speculative world grounded in the future consumption of biohybrid flying robots.

The visual identity of the event was themed around the color blue, \zm{for its associations with the} sky, freedom, and artificiality. This motif extended from the lighting and interior setup to drinks and desserts, invoking a sense of alien futurity. Table settings included napkins (for the Ortolan ritual), feathers, bird dolls, and glassware shaped like birds. These theatrical elements were chosen for their symbolic and sensory resonance, blending natural motifs with synthetic speculation. All guests were instructed to place napkins over their heads during the main course, reviving the traditional Ortolan ritual in a new technological context.

The atmosphere was further enriched through sonic design. At different narrative moments, we used the nostalgic French song “L'amour est bleu,” authentic Ortolan birdsong, and artificially generated digital bird melodies to mark the emotional and ontological shifts across the experience—from nature to machine, from real to speculative.

We created and deployed three primary props (see \autoref{fig:props}), each representing the biohybrid robot in a different stage of its imagined lifecycle:

\begin{itemize}
    \item \textbf{Prop A – Flying Drone:} A feather-adorned quadcopter was flown through the dining space via Wizard-of-Oz techniques \cite{woz}. The drone’s dynamic flight conveyed a lifelike presence and introduced movement into the sensory landscape.
    
    \item \textbf{Prop B – 3D-Printed Prototype:} Created from translucent resin, this prototype abstractly depicted the internal anatomy of the hybrid: colorful bones (bright blue), visible yellow fat, semi-transparent skin, and a geometric head unit to suggest a camera-driven AI controller. The prop was designed to bridge biological and mechanical aesthetics, making the hybrid’s fictional anatomy visible and plausible.
    
    \item \textbf{Prop C – Plated Main Course:} A roasted quail represented the edible, final form of the hybrid, directly referencing the Ortolan in size and taste. It served to ground the speculative fiction in culinary realism and highlight ethical substitution.
\end{itemize}

\textbf{Design Rationale and Iteration:} The design process for the 3D-printed prototype (Prop B) evolved through several iterations. Inspired by anatomical diagrams, speculative fiction aesthetics, and edible robotics research, we gradually refined the head from a naturalistic form to a trapezoidal shape suggesting both abstraction and embedded intelligence. The translucent resin allowed participants to "see inside" the body, metaphorically suggesting transparency in future food production. While we did not involve bioengineering experts in prototyping due to resource constraints, we drew from interdisciplinary inspirations to ensure aesthetic coherence and thematic provocation.

\textbf{Culinary Design and Symbolism:} The three-course menu (see \autoref{fig:menu}) reflected the narrative arc of the drama and incorporated speculative and symbolic ingredients. The starter mimicked the natural diet of Ortolans, placing participants in the imagined perspective of the bird. The main course substituted Ortolan with quail for ethical, legal, and sensory alignment, allowing diners to experience a taste analogous to the original dish \cite{poon_2014}. The final dessert—a futuristic blue concoction—reinforced the speculative setting and closed the meal with a sense of surrealism. Real ingredients were disclosed in advance to respect dietary needs and ethical transparency.

\textbf{Research Role of Prototypes:} While these were performative props, they functioned as epistemic artifacts—helping to make the speculative scenario legible, thinkable, and emotionally engaging. Their design was not solely aesthetic but methodological: each embodied a phase in the ontological ambiguity of life, machine, and food, triggering participant reflections central to our research questions.

\zm{Further details of the visual and morphological design process, including intermediary sketches and iterative explorations of wings, head structures, and textural elements, are provided in ~\autoref{sec:design_appendix}. These materials complement the performative account presented here by documenting the broader development of the fictional biohybrid creature.}

\subsection{Participants and Their Role Play}\label{participant}

We recruited six guest participants using snowball sampling, all professionals working in creative, design, and cultural sectors. This \zm{sampling} was intentional: \zm{creative professionals often serve as “lead users”} with imaginative capacities and critical reflection skills to engage productively in speculative inquiry. The participants were: P1, a design lecturer at a world-renowned institution; P2, an award-winning poet; P3, an artist and gallery owner; P4, a foresight expert from a global tech innovation hub; P5, a professional UX designer; and P6, a filmmaker. The group was gender-balanced,  with five participants in their 30s and one in his 70s. Participants were verified as meat eaters, and food allergies were checked in advance. One participant with dietary restrictions participated as a non-eater, taking on the role of the restaurant service manager.

The event was framed as a ``dinner-in-the-drama'' taking place in a fictional 2052 Paris restaurant. Participants were invited to inhabit customized personas inspired by their real-world occupations, allowing for semi-structured but naturalistic role-play. Each persona was paired with discussion prompts thematically aligned with their real-life expertise (e.g., environmental, ethical, political). This mapping sought to support ecological validity, encouraging authentic engagement within a speculative fiction setting while reducing performance pressure or role incongruence.

While role-playing introduced fictional framing, participants were encouraged to speak from their own perspectives, enabling a hybrid mode between improvisational acting and real-world reflection. This method allowed us to observe how participants explored, negotiated, and responded to speculative concepts, while remaining anchored in their lived knowledge and values. Furthermore, assigning a range of perspectives ensured balanced discussion and mitigated dominant voices. \autoref{tab:p} demonstrates the participants' roles, assigned personas, and discussion topics. 

All participants provided informed consent to be audio- and video-recorded, photographed, and analyzed. They were told the session was part of a research project and were free to withdraw at any time. As compensation, food and event-related costs were covered, and participants enjoyed the dinner as a unique speculative experience.

\subsection{Event Procedure}
\begin{figure*}
    \centering
    \includegraphics[width=\linewidth]{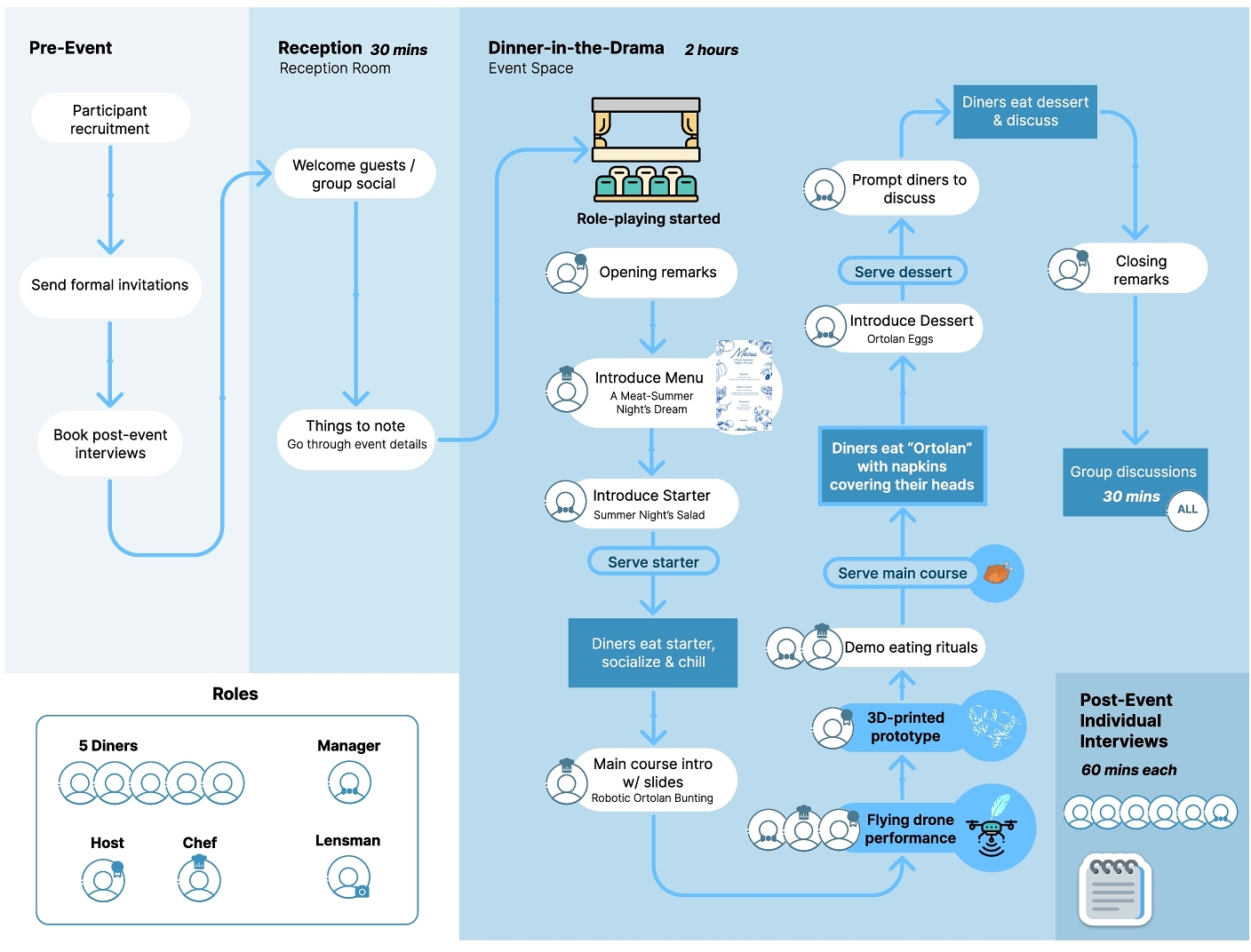} 
    \caption{Procedure diagram.} 
    \Description{A nicely illustrated flowchart.}
    \label{fig:prcd}
\end{figure*}
To explore how people might engage with the idea of consuming biohybrid animal-robot hybrids in the future, we designed a dinner-in-the-drama event that unfolded across multiple scripted and improvised stages (\autoref{fig:prcd}). The flow and content of the event were deliberately crafted to elicit both emotional and reflective responses around the speculative scenario.

Upon arrival, participants attended a reception with drinks and snacks to foster rapport and comfort. They were then ushered into the dining room, where a staged environment evoked the future 2052 Paris restaurant. Cultural cues---such as the projection of the Eiffel Tower and the soundtrack “L'amour est bleu”---anchored the experience in both a futuristic and distinctly French atmosphere.

\begin{figure*}
    \centering
    \includegraphics[width=\linewidth]{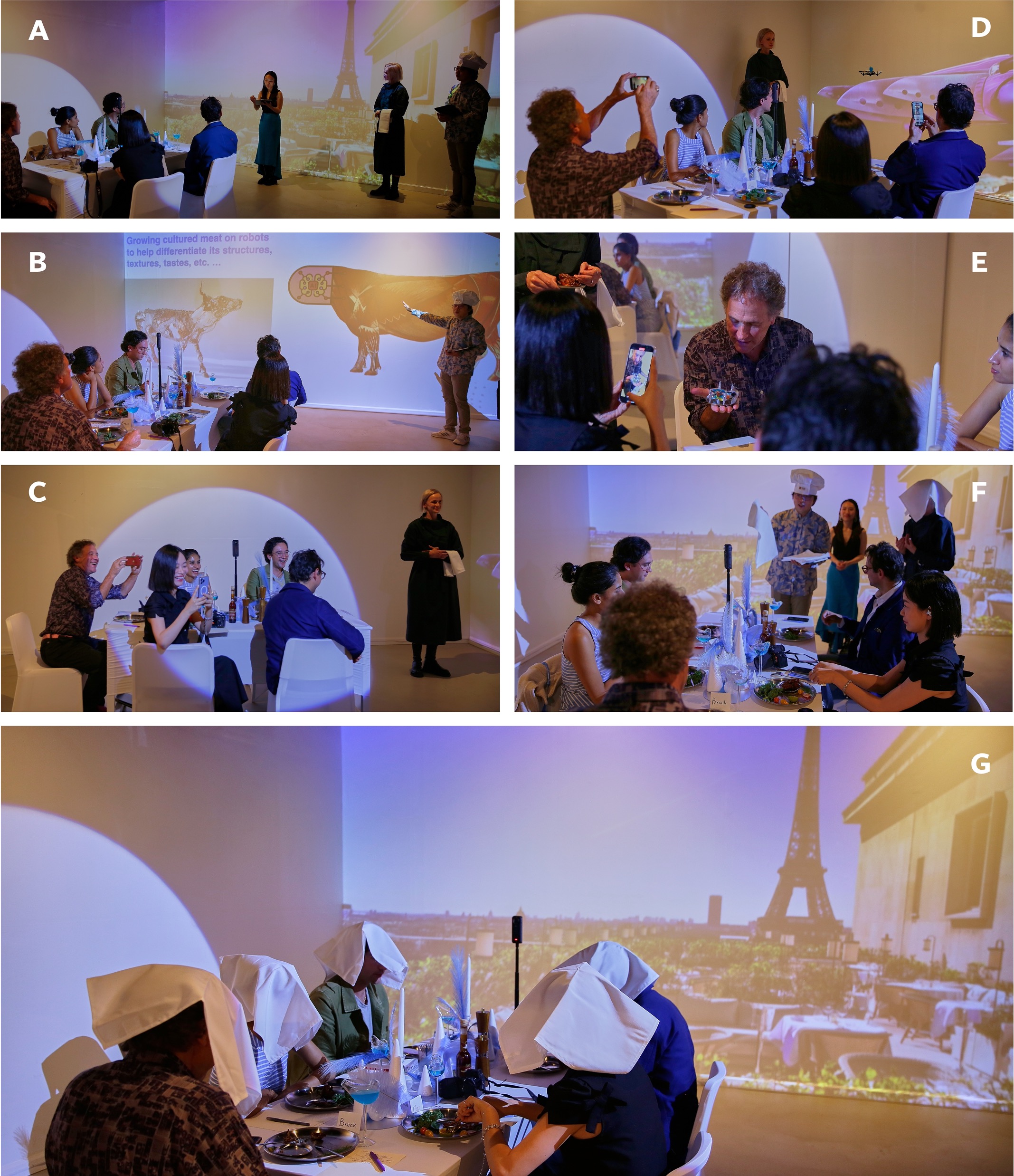} 
    \Description{Different moments of the event that a group of people were participating.}
    \caption{Highlights from the dinner-in-the-drama: (a) Opening remarks; (b) Introducing key concepts; (c) Enjoying the show; (d) Biohybrid robot in flight; (e) Observing the biohybrid robot; (f) Serving biohybrid robots; (g) Eating biohybrid robots.} 
    \label{fig:event}
\end{figure*}

The host (lead artist) welcomed guests and introduced the speculative narrative. The chef (lead researcher) then explained the fictional menu and contextualized the key concept of edible animal-robot hybrids. Dramatic beats followed: a feathered drone representing the “synthetic ortolan” was flown across the room (\autoref{fig:event}d), after which it was replaced by a 3D prototype passed among the guests (\autoref{fig:event}e). The final edible course---a roasted quail representing the hybrid---was served under napkins, referencing the historic Ortolan tradition (\autoref{fig:event}f). \zm{This sequence was designed to gradually increase sensory immersion, shifting participants from observational to embodied engagement, and culminating in reflective dialogue.} The experience concluded with a dessert and a moderated post-dinner discussion. \zm{The event was video-recorded using a 360° omnidirectional camera, and the researchers took field notes throughout the session.}

\subsection{Post-Event Interviews and Data Analysis}
After the immersive dining event, semi-structured individual interviews were conducted with each participant. The interview questions aimed to gather participants' reflections on their overall experience of the drama and dinner event, with a focus on both the sensory and ethical dimensions of the staged future. Participants were asked to share their thoughts on the drama itself and provide opinions on key topics, including philosophical, ethical, environmental, economic, and socio-cultural issues from PESTEL Framework \cite{yuksel2012developing}. Additionally, we explored their perceptions of the identity of the animal-robot hybrid and their views on AI-infused food systems. \zm{Interview questions are listed in \autoref{sec:interview}.}

\zm{Interviews were audio-recorded and transcribed. Given the study’s small and exploratory sample, we employed an inductive thematic analysis approach \cite{braun2012thematic} to surface rich, situated insights. Two researchers independently familiarized themselves with the transcripts and generated initial codes, ensuring analytical depth and reducing individual bias. The researchers then met to compare, refine, and consolidate codes into preliminary themes, resolving discrepancies through discussion and returning to the data as needed. Emerging themes were iteratively cross-checked against field notes and observational data from the event to strengthen interpretive validity and maintain alignment with participants’ embodied experiences. This reflexive, multi-stage process supported a trustworthy and transparent interpretation of the qualitative material.} The goal of this study was not to generate generalizable claims, but to probe diverse reactions and reflections using a depth-over-breadth approach common in design research and speculative methods.


\section{Results}\label{results}
Overall, participants expressed curiosity, caution, and concern about the role of biohybrid robotics in reimagining food. In our critical exploration of alternative food futures (see \ref{altfoodfut}), we identified three themes: \textit{Technological Feasibility and Pathways; Ethics and Moral Boundaries; and Symbolism, Culture, and Rituals}. Each theme integrates multiple perspectives, capturing both convergence and divergence. We also examine how participants engaged with the research method itself (see \ref{tanmuldin}) through three lenses: \textit{Immersive Engagement and Emotive Responses; Multisensory Cross-Media Theatrical Design; and The Serendipity of Unexpectedness and Improvisation}.

The reflections presented below are not offered as authoritative predictions or ethically prescriptive positions. Nor do they represent a unified vision for the aspirations surrounding emerging food technologies. Rather, they represent situated perspectives that illuminate a contested terrain where hopes, anxieties, and imaginaries of possible food futures---biohybrid or otherwise---circulate and collide.

\subsection{Critical Explorations of Alternative Food Futures}\label{altfoodfut}

\subsubsection{\textbf{Technological Feasibility and Pathways}}\label{techfea}
Participants offered varied yet complementary perspectives on the maturity, timelines, and scalability of biohybrid food technologies. Several viewed lab-grown meat and partial integrations—such as meat grown on robotic substrates—as plausible in the near term, while fully integrated biohybrid robots producing edible tissue were imagined as a more distant possibility. Additionally, while some envisioned synthetic food technologies catering to niche markets, others anticipated broader economic implications, including challenges related to affordability and market accessibility.

P6 suggested that building an autonomous biohybrid animal with an AI “brain” would be prohibitively costly but not inconceivable, estimating a 40-year horizon before such systems might emerge. In the meantime, partial implementations are likely to emerge, serving as marketing tools to drive early adoption. By contrast, P3 drew on science fiction imaginaries, envisioning similar futures within a decade but stressing that adoption should be slowed to allow society to grapple with social and ethical complexities. 

P4 highlighted taste as an immediate and sensorial barrier to adoption, arguing that improvements in flavor would accelerate acceptance and enable technological scaling. He suggested that market pressures could drive innovation toward technical customization---moving beyond familiar forms toward modular systems that allow tuning of flavor and nutritional profiles. This could involve experimenting with novel material combinations and biological processes, including cross-species or hybrid formations (e.g., “tiger eggs”), potentially supported by AI to optimize taste, umami, and nutrient composition.

P4 also envisioned a staged diffusion despite high costs, \textit{“Early adopters will…try it anyway…then the fast follower markets like experiencers who would do it for the thrill…whereas the more information-seeking market…will be more interested in sustainability, environmental impacts…[and] whether it fits into their lifestyle.”} Complementing this view, P2 invoked a “Tesla model” of innovation, in which early, expensive applications enable the gradual maturation of supply chains and cost reductions.

Beyond consumer markets, participants speculated on system-level effects. P4 pointed to food supply chains and the political stakes for governments choosing between promoting innovation or protecting traditional farming. Regarding edible biohybrid robots, participants envisioned a dichotomy between homogenized mass production and artisanal, bespoke biohybrids with individual identities. This distinction structures divergent ethical and cultural consequences, which are examined in the following subsections and synthesized in \autoref{tab:indihomo}.
Extending this vision, P6 imagined customization at the production level, with ranchers acting as curators who fine-tune traits such as personality and taste to create differentiated offerings.

Participants also questioned whether synthetic food production is as environmentally sustainable as often claimed, particularly when considering its ecological impacts relative to other alternatives. Concerns included the resource intensity of engineering food at scale (P1), the energy demands of biohybrid and AI-driven processes (P2, P5), and the broader ecological footprint relative to traditional farming (P2). For instance, P2 criticized the current efficiency of lab-grown meat, stating that \textit{“It takes tons and tons of energy to do badly something that you'd use much less energy just raising a cow.”} P5 extended the discussion by suggesting that understanding our relationship with animal–robot hybrids may be more important than their potential environmental impact.

\begin{table*}[h]
\caption{Individuality vs. Homogeneity in Biohybrid Food Futures}
\label{tab:indihomo}
\centering
\begin{tabular}{p{0.2\textwidth} p{0.37\textwidth} p{0.37\textwidth}} 
\toprule
\textbf{Dimension} & \textbf{Individuality} & \textbf{Homogeneity} \\
\midrule
\textbf{Core Form} & Differentiated, customized, small-scale. & Standardized, indistinguishable, mass-produced. \\
\textbf{Market Analogy} & Small farms or locally sourced meat. & Factory farming and industrial meat production. \\
\textbf{Market Position} & Premium, bespoke products. & Affordable, widely accessible products. \\
\textbf{Governance} & Contentious, politically sensitive regulation. & Compatibility with existing regulatory frameworks. \\
\textbf{Human Perception} & Heightened lifelikeness; blurred animal–machine boundaries. & Machine-like; reinforced technological status. \\
\textbf{Ethical Implications} & \textbf{Heightened ethical tension}: perceived personality elicits empathy, complicating harm or consumption. & \textbf{Reduced ethical tension}: commodification eases justification and regulation. \\
\textbf{Food Culture} & Ritualized, sensory, affective engagement. & Utilitarian, routine consumption. \\
\bottomrule
\end{tabular}
\end{table*}

\subsubsection{\textbf{Ethics and Moral Boundaries}}
Ethical considerations emerged as a central axis of reflection, ranging from debates about sentience and relativist ethics to extreme speculative scenarios. Participants posed ethical tests and drew on imaginaries from adjacent fields such as synthetic biology, robotics, and AI. Across these discussions, they wrestled with how biohybrids might provoke new moral dilemmas, situating edible biohybrid robots within broader debates about technological agency, taboos, and moral boundaries.

\textbf{Sentience and Suffering.}
P6 observed that humans routinely anthropomorphize machines and animals, projecting sentience and emotion onto them. For him, this raised a central objection: \textit{“We would be building a sentient being to kill it… there would be more public resistance.”} P1 extended this: 
\begin{displayquote}
\textit{“If we are making these things more similar to living organisms… at what point do they just approach the same organisms we are trying to replace and prevent suffering?”}
\end{displayquote}

P3 broadened the debate by invoking religious worldviews in which all living things possess a soul; she extended this to AI-driven or biohybrid systems: \textit{“Could they embody a form of soul?”} P2 noted that resolving such questions requires \textit{“a philosophical position on the soul and materialism”} to judge whether biohybrids might possess sentience and thus the capacity to suffer.

Acknowledging the intractability of such ontological debates, P2 offered an ethical test: if a biohybrid \textit{“looks like an animal”} but \textit{“does not have feelings”}, harming it would be \textit{“ethically neutral”}. He extended this provocatively, \textit{“Is it unethical to grow a human and eat the human? It’s not a feeling person… there’s no ethical harm.”} 
While framed hypothetically, participants reacted with visible discomfort. P2 himself acknowledged the idea as \textit{“powerful and gross”}, revealing the gap between harm-based ethical reasoning and deeply embedded social taboos that function as moral boundaries.

P2 further invoked \textit{Westworld} and \textit{Ex Machina}, noting how popular debates about machine ethics focus on humanoid robots while largely overlooking animal-like entities and biohybrids. In contrast, he emphasized that the \textit{``particular design [of the animal-robot hybrids] brought the conversation into the realm''}.

\textbf{Relativist Ethics and Tradeoffs.}\label{Ethics}
Participants reflected a relativist ethical stance in which the moral evaluation of biohybrid animals depends on context, form, and social framing rather than fixed principles. For example, P5 recounted a story of a grandmother in rural Alaska who killed and cooked a cat after it ate rare cheese, underscoring how scarcity and circumstance reshape moral boundaries: \textit{“It’s the context and the living environment where you are…for them it was survival.”} This perspective extended to synthetic foods and biohybrids, where specific situations complicate simple distinctions between food and companion: \textit{“Do you think about it as food or do you think about it as an animal—why can’t it be both?”}

P6 drew an analogy between polarized views on factory farming and locally sourced organic meat, suggesting that synthetic food could be framed along similar ideological lines, see \autoref{tab:indihomo}. He expressed a preference for individualized biohybrid animals—akin to small-farm meat—arguing that their distinct traits and personalities heightened ethical sensitivity and made the experience more impactful: \textit{``If all of these animals had their own personalities, then people seem to be more wary.''} P5 echoed that such individuality invites people to perceive biohybrids as subjects rather than interchangeable objects, fostering empathy, attachment, and a sense of moral responsibility that complicates purely instrumental or efficiency-driven reasoning around harm or consumption. By contrast, more homogeneous biohybrids were seen as eliciting less concern, as standardization and replaceability allow them to be framed as commodities or infrastructure, making large-scale production easier to justify. 

P1 reflected on how ethical production practices might unintentionally diminish flavor, citing hothouse tomatoes as an example. She noted that tomatoes grown in stress-free greenhouse conditions often taste bland, whereas the sweetness and complexity of Mediterranean tomatoes emerge from environmental stressors such as uneven sunlight and pest exposure. This led her to question whether eliminating suffering—whether in plants or animals—might also erase some of the sensory richness people value, revealing a tension between ethical responsibility and culinary pleasure. Extending this idea, P6 described rare or unusual foods—such as bird’s nest soup, shark fin, or ortolan bunting—as “special” precisely because of their rarity and infrequent consumption, suggesting that novelty and scarcity can heighten sensory and emotional value even as ethical concerns remain salient.

\textbf{Replicability and Runaway Risks.} P1 warned that creating self-replicating biohybrids could introduce runaway effects, considering the idea of AI-led integration of synthetic and biological structures in robotics. For P1, the risks were profound: \textit{“Life is a machine for self-perpetuation, so the odds of unintended consequences are so much higher for creating a living agent.”} If capable of replication, biohybrids would enter \textit{“the category of any living being with an intrinsic drive to reproduce”}, potentially acting like a virus against other systems. 

These anxieties connected to wider fears about AI, where distributed agency and emergent behavior defy governance. As P1 put it, \textit{“It connects to AI… fears about general artificial intelligence having runaway impacts.”} Such reflections were framed as calls for humility, guardrails, and continuous evaluation. P5 echoed this, questioning: 

\begin{displayquote}
\textit{“How much control humans should exert: should biohybrids have feelings, agency, or senses? How much autonomy should they be allowed to develop?”}
\end{displayquote}

\subsubsection{\nb{\textbf{Symbolism, Culture, and Rituals}}}

Participants noted that the symbolic dimensions of food technology shape both the preservation and transformation of culture and rituals. P1 emphasized familiarity as a bridge to help people adjust to innovation, expressing hope that food continues to carry cultural meaning rather than serving purely as sustenance. She argued that biohybrids would initially adopt the \textit{“trappings of traditional food”} to ease novelty acceptance; over time, however, biohybrids would move beyond imitation, giving rise to new aesthetics, flavors, and rituals. Similarly, P2 referenced the principle of MAYA (\textit{Most Advanced Yet Acceptable}) to stress that designs balancing familiarity and novelty can smooth transitions from niche enthusiasm to mainstream uptake.

Extending this discussion, P2 commented that the dinner theater revealed \textit{“a whole dimension about culture, cultural preservation, and cultural innovation”}, describing the creation of the ortolan robot as \textit{“a positive and clever way to preserve a tricky piece of culture—using technology to ethically achieve what was once unethical or has since become problematic”}. He further reflected on how cultural traditions can be preserved and ritual meaning maintained even as material forms change. He noted that religious and ceremonial diets have historically adapted to new conditions while retaining symbolic resonance, such as substituting animal sacrifice with ritual acts using bread. Through such adaptations, he argued, societies can sustain the meaningful dimensions of ritual while accommodating ethical, technological, or practical constraints.

In contrast, P5 noted, \textit{“Beef is such a politically charged topic in India because cows are sacred to Hindus.”} She speculated that lab-grown beef could, in principle, enable those who wish to eat beef to do so while respecting those who oppose killing cows. Yet she doubted whether technological innovation alone could bypass such deeply rooted sensitivities, asking \textit{“If [the cow is] lab-grown… does it solve all our problems?”} As she argued, cultural meaning often outweighs materiality:

\begin{displayquote}
\textit{“Sometimes it doesn’t matter whether it’s real or synthetic, it’s just the fact that this is a symbol of my faith or my beliefs.”}
\end{displayquote}

P5 also reflected on geographic diversity, pointing out that culinary innovation is often associated with specific locales—like Paris as a center of gastronomy, but noted that \textit{“other cultures with rich culinary history”}, such as China or India, could equally serve as hubs for future food innovation, underscoring the interplay of local tradition, ethical concerns, and global imagination in shaping the future of eating.

Participants highlighted how biohybrid animals and synthetic meat can serve as media for cultural and artistic expression and as catalysts for cultural innovation. Distinct individuality in biohybrids was seen to enhance narrative engagement and provoke ethical reflection among diners. P6 suggested that by integrating the life stories and distinctive traits of biohybrids into dining experiences, restaurants could offer richer storytelling that engages both emotion and morality. Extending this idea to the domestic sphere, P5 envisioned personalization through family-led assembly or breeding of unique variants, producing heirloom-like recipes passed down across generations and fostering affective attachments.

\subsection{\nb{Tangible Multisensory Dining Experience as a Valuable Provocation}}\label{tanmuldin}

As shown above, our method and multisensory design surfaced a rich and productive range of perspectives and tensions. Participants consistently connected their technological speculations to deeper reflections on politics, religion, nationality, literary canons, symbolism, and ontological beliefs about the nature of life, consciousness, and suffering. In what follows, we explore three mechanisms through which these deep speculative exchanges emerged. 

\subsubsection{\textbf{Immersive Engagement and Emotive Responses}}

Participants felt that the immersive dining format---particularly the use of napkin hoods and live/mechanical elements---provoked strong sensory engagement and elicited intense emotional responses. Several highlighted the hooded moment as especially impactful, though their interpretations diverged.
P2 described how covering heads with napkins transformed the room: conversation ceased and a collective silence emerged, \textit{“like putting a cover over a birdcage”}. Rather than diminishing sociality, this paradoxically heightened his awareness of others: he became \textit{“hyper-aware of the other people eating around me”}, sensing a subtle connection through shared quietude. By contrast, P4 found the same setup isolating and unsettling, likening the stillness to the loneliness of COVID-19 lockdowns.

P5 found the hood sharpened perception of detail, including the textures and smells of food, \textit{“It does help you focus on that, like this is what you’re eating.”} P6 appreciated the privacy, \textit{“I didn’t really like…watching you eat bones… so it was nice to eat freely and not feel like people were watching me… it made the experience more intimate and helped me really focus on the act of eating.”}
P3 described the most ambivalent reaction: feelings of \textit{“shame and guilt”} as if engaging in something forbidden, mixed with curiosity and excitement. Crucially, she stressed that these tensions stemmed less from the food itself than from the performative act of concealment. For her, hiding under a napkin generated novelty, surprise, and playfulness alongside discomfort.

Overall, these divergent responses show how ritualized interventions can shift social dynamics and reshape sensory perception, interpersonal awareness, and ethical imagination, provoking new avenues for reflection.

\subsubsection{\textbf{Multisensory Cross-Media Theatrical Design}}


Participants reported that cross-media representations, including slideshow drawings, decorative birds, the live drone, the 3D-printed model, and the edible roasted birds shaped their perception of the boundary between “animal” and “machine”.

Design cues strongly mediated anthropomorphism and engagement: P5 observed that the absence of eyes in the model helped it remain object-like: \textit{“The moment you have eyes, I think, makes it a person… the absence of that helped.”} P1 likened the 3D print to a “butchered” version of the bird-drone, highlighting contrasts between living and mechanical forms. P4 emphasized the robot-like qualities of the hybrid, noting its mechanical parts and the design logic of starting with a robot and adding biological material; he noted how gradual addition of human features can shift perception but preferred the hybrid to remain clearly machine-like, reinforcing its role as a meat-production device. 

P2 noted that the cross-media layering representations suggested possibilities for future iterations, imagining a synthesis of form and function: \textit{“I could imagine… with more resources… to have the drone technology that makes it function join with that beautiful 3D printed resin model.”} He further suggested that adding feathery textures on the drone model could enhance perceived vitality, noting that even the small decorative birds (see \autoref{fig:details_appendix}b) on the table contributed a stronger sense of life and observation. Participants also noted that seeing slideshow representations (e.g., \autoref{fig:event}b) of how animals are raised, prepared, and consumed clarified both practical and ethical aspects.

The multisensory hands-on, participatory character of the theatrical method was also generative. P6 described handling the models and watching the drone as a form of imaginative play: \textit{“I’m totally game for that… I can pretend this is a bird and not a drone… it led to a good moment where the drone kind of had a mind of its own.”} P4 and P5 similarly emphasized that physically interacting with the models and drones stimulated richer imagination and reflection than written reports or static visuals. P2 extended this idea, framing the dining itself as a medium of engagement: \textit{“We were literally putting the meat from the presentation inside our bodies”}---making imagined futures both tangible and memorable. P3 added that \textit{“It felt like we were acting in a sci-fi movie.”} As P4 summarized, such interactive experiences are \textit{“much more effective”}, while P2 went further, positioning play as the key mechanism:
\begin{displayquote}
 \textit{``This is a wonderful way to help people engage with a speculative future... there was a situation that allowed us all to dwell and process in it.''}
\end{displayquote}




\subsubsection{\textbf{The Serendipity of Unexpectedness and Improvisation}}

All participants noted that an unexpected drone malfunction profoundly shaped their experience of the drama and provoked deeper reflections. During the demonstration, insufficient ambient lighting caused the drone’s vision system to fail, sending it off course and into the corner of the room. This brief “out-of-control” moment became a focal point of the experience; \autoref{fig:event}c captures participants’ reactions.

\begin{displayquote}
\textit{“I especially like the part where it tried to escape... well, it made it more real... it's trying to get away.”} (P4)
\vspace{2mm}

\textit{“That made it much stronger for the guests and much more thought-provoking because then this thing had its own free will.”} (P1)
\vspace{2mm}

\textit{“It reminded us that machines sometimes malfunction and that humans aren't always perfect at making things go how we want with our technology.”} (P2)
\vspace{1mm}
\end{displayquote}

Similarly, P5 described the erratic drone behaviour as a \textit{“reminder that speculative technologies will inevitably go wrong”}, emphasizing the importance of anticipating risks and strategies for response. For P3, the same incident was striking, describing the crash as \textit{“really beautiful”}, suggesting that the uncontrolled movements imbued the machine with \textit{“a sense of soul”}. Such unexpected moments can be seen to directly inform the generation of substantive themes unpacked in \ref{Ethics}. 

While drone malfunctions introduced unexpected moments into the drama, equally significant were participants’ unplanned contributions as they role-played and interacted with one another. Group discussion was encouraged throughout and after the curated experience. P2 reflected that the act of eating together also provided a conducive social environment for exchanging viewpoints and speculation, creating an imaginative setting that an individual alone could not achieve. This social environment, full of fast-paced lively exchanges, encroached on territories that would unlikely be given thought through individual contemplation.

Participants were invited to improvise with the narrative, props, and environment of the design fiction. This opened space for idiosyncratic references to literature, philosophy, and popular culture to surface. For example, P2 connected the concept of creating biohybrid robot meat to Shakespeare’s line, \textit{“the green-eyed monster which doth mock the meat it feeds on”}, commenting that the dinner theater was effectively \textit{“mocking meat”}. He used the pun to suggest that the edible robot enactment both imitated natural meat and satirized the problems of contemporary meat production. This dual meaning highlights a paradox: alternatives designed to be ethically superior may still mimic the very practices they aim to replace. 

Together, these accounts frame unexpectedness and improvisation not as disruption but as a speculative and provocative resource. While such moments cannot be fully designed or foreseen, the open, immersive, and performative nature of the dinner-in-the-drama created conditions where this productive serendipity could unfold---revealing how surprise, malfunction, and improvisation can deepen reflection on technological ethics and imagined futures.

\section{\zm{Discussion}}\label{discussion}
\subsection{Ambiguity in Lay Imaginaries of Biohybrid Foods}
Participants’ future imaginaries of biohybrid foods were dynamic, oscillating, and sometimes contradictory. They speculated about timelines, market adoption, taste, sustainability, and social impacts, revealing a complex landscape of hopes, anxieties, and ethical considerations. Rather than treating these oscillations as misunderstandings or gaps in knowledge, we interpret them as reflections of the inherent ambiguity of biohybrid foods. Prior speculative proposals around robot meat and edible robotic systems suggest that this ambiguity is not unique to our study but characteristic of broader imaginaries of hybridized food–machine entities \cite{robotmeat, floreano2024edible, zhang2025edible}. Drawing on Gaver’s concept of ambiguity as a resource \cite{Gaver} and Barad’s notion of indeterminacy \cite{barad_2007}, we conceptualize biohybrid foods as inherently ambiguous technologies: entities that actively blur boundaries between biological and mechanical, alive and designed, edible and symbolic.

The speculative dining format amplified this ambiguity by foregrounding sensory and embodied provocation. Taste, texture, sound, and ingestion rituals encouraged participants to model critical scenarios of adoption, asking how biohybrids might enter everyday life, be regulated, or be personalized. This embodied engagement enabled participants to explore future imaginaries and ritual transformation, imagining how culinary practices, cultural rituals, and ethical norms might shift in response to hybridized foods.

At the same time, several forms of ambiguity emerged through unscripted and unexpected moments during the dining encounters. These moments were not designed as fixed narrative prompts; they arose contingently through sensory surprise, improvisation, or unplanned technical errors. In line with arguments that ambiguity invites multiple interpretations and sustained personal engagement \cite{Gaver}, these unpredictable encounters prevented any single interpretive frame from stabilizing. Such “productive indeterminacy” is not a deficit of knowledge to be resolved, but an ongoing material–discursive process in which boundaries come to matter through practice \cite{barad_2007}. In this sense, ambiguity was not merely present but intensified through emergent, unanticipated interactions with speculative artifacts.

The oscillating and situated nature of these imaginaries shows that ambiguity is not a barrier to understanding; it is a productive epistemic and design space that fosters ethical imagination, collective sense-making, and culturally grounded speculation. Through this lens, we contribute to work in HCI, HRI, and HFI by identifying ambiguity itself as an empirical pattern in lay imaginaries of emerging food technologies. Rather than treating oscillation as noise or confusion, we show it to be a structuring feature of how people make sense of biohybrid futures through sensory, ethical, and affective reasoning. This subsection therefore offers a conceptual shift: indeterminacy is not simply a theoretical property of hybrid organisms, but a situated cognitive and social practice enacted through engagement with ambiguous artifacts \cite{Gaver, barad_2007}.

\subsection{Edibility as a Negotiated Ethical Boundary}
Participants’ encounters highlighted edibility as a central and negotiated ethical boundary. Rather than a fixed property, edibility emerged as a boundary object connecting biology, technology, culture, and personal experience \cite{star_griesemer_1989, probyn2016eating, Paxson, bingham_2006}. Questions surfaced regarding conditions under which consuming hybrid organisms might be acceptable, and what forms of responsibility are owed to semi-living entities. These concerns echo ongoing debates in biohybrid robotics ethics, where researchers argue that boundaries between organism, technology, and moral obligation are negotiated rather than given \cite{biohybrid_ethics, kirksey2021living}. As others note, imaginaries of “living machines” often draw from cultural narratives of control over life \cite{vaage}, and here such narratives are refracted through the intimate act of ingestion.

Ethical reflection was embodied and relational. Taste, texture, and movement shaped judgments that oscillated between attraction and aversion, curiosity and caution. Participants invoked frames from synthetic biology, AI ethics, and religious belief \cite{hyun2016embryology, internationalsocietyforstemcellresearch_2021, Webster-Wood_2023} to reason about sentience, individuality, and replicability. In this way, biohybrid foods prompted situated renegotiation of “edible versus inedible”, making moral and cultural boundaries explicit and mutable. Viewed through posthumanist perspectives \cite{braidotti_2013, donnaharaway_2008}, these reflections foreground human–nonhuman entanglements, where ethical reasoning arises through encounter with semi-living, partially autonomous entities rather than abstract principle.

Speculative dining functioned as a method for ethical boundary-work, enabling participants to experiment with norms, values, and social practices around hybrid foods. Instead of resolving whether biohybrids should be eaten, the dining encounters surfaced how technological, sensory, cultural, and ethical considerations co-produce what counts as food, and under what conditions. For HCI and HFI, this positions edibility not as an inherent attribute of objects, but as a relational and negotiated property continually shaped through sensory interaction and shared deliberation.

\subsection{The Cultural and Affective Politics of Synthetic Eating}
Participants’ reactions to the speculative dishes demonstrated that synthetic eating is not merely an ethical or sensory practice, but deeply embedded in cultural values and affective frameworks. Their interpretations invoked memories of family traditions, religious norms, regional culinary identities, and deeply held taboos---showing that biohybrid foods enter cultural fields already structured by history, identity, and emotional attachments. As previous work in food politics argues, eating practices are inseparable from collective identity, moral order, and social meaning \cite{counihan_2013, fischler_1988}, and our results confirm that participants mobilized such frameworks to make sense of hybridized foods.

Throughout the dining sequence, individuals continuously mapped unfamiliar specimens onto familiar cultural categories, echoing anthropological accounts of how novel or “exotic” foods are domesticated through analogy and metaphor \cite{evan_sutton_2006, fischler_1988}. These mappings were not superficial analogies: they materially shaped participants’ willingness to engage, emotional responses, and their narratives about potential futures. For example, when a hybrid specimen was associated with an existing ritual food, some participants considered it acceptable for ceremonial consumption; when another evoked tabooed species, unease or refusal followed. This pattern aligns with research in Human–Food Interaction showing that sensory and cultural cues (e.g. smell, texture, appearance) significantly influence the perceived acceptability of unfamiliar food–technology hybrids \cite{obrist14, obr16}.

Beyond cognitive mapping, the experience triggered a spectrum of affective responses---curiosity at novelty, discomfort at blurred organism–machine boundaries, protectiveness toward malfunctioning drones or “fragile” tissue samples, and even excitement about future possibilities. These emotional reactions played a structuring role: discomfort often led to rejection, curiosity prompted further engagement, and care sometimes reframed hybrids as quasi-relational entities deserving ethical consideration. This resonates with scholarship that positions emotion and affect not as background reactions but as central in stabilizing---or destabilizing---categories of edibility, purity, and moral acceptability \cite{probyn2016eating, fischler_1988}. In our context, affective engagement was facilitated through material and interactive design---for instance, by invoking uncertainty, fragility, or familiarity through multisensory artifacts, akin to how edible-robotic or microbial food-technology work uses sensory design to elicit intimacy, care, or repulsion \cite{floreano2024edible, Chen2021}.

Participants also frequently invoked broader social norms and anticipated public reactions---imagining how others might respond, speculating on collective acceptance or resistance, and foreseeing that hybrid foods could trigger new rituals or social divisions. This echoes classic anthropological theory that food practices act as cultural boundary markers \cite{douglas_1972}. In our study, synthetic eating triggered three types of boundary-work: (i) reinforcing existing cultural distinctions (e.g. natural vs. unnatural, sacred vs. profane), (ii) reworking traditions by imagining hybrid meals that combine conventional and novel elements, and (iii) projecting future cultural orders in which hybrid foods serve as markers of ecological adaptation, social status, or technological modernity. Similar projections appear in speculative design and food-futures literature, where imagined consumption practices become sites of social imaginaries and value negotiation \cite{dunne2024speculative, participatefood}.

Together, these findings highlight that adopting biohybrid foods cannot be understood solely as a question of technical feasibility or individual taste---it is fundamentally a process negotiated through cultural frameworks, affective attachments, and imagined social futures. For HCI, HFI, and HRI research concerned with emerging food technologies, these results underscore the importance of attending not only to design and usability, but also to the cultural infrastructures and affective economies through which hybrid foods may acquire meaning and legitimacy.

\subsection{Tangible Design Fiction as Ethical Sandbox and Public Co-Theorization}

Speculative dining functioned as a tangible form of design fiction in which ethical and ontological questions were enacted through ritualized, multisensory practice rather than merely discussed in the abstract. Hooding, cross-media staging, and direct engagement with hybrid food objects generated designed disruptions---moments of curiosity, discomfort, and collective attentiveness that foregrounded shifting relations of visibility, agency, and responsibility \cite{auger_2013, michael_2012, jewitt_barker_steimle_2022}. These interventions did not simply represent imagined futures; they staged conditions for experimenting with ethical boundaries in situ. This extends prior work on speculative design showing that embodied, non-representational encounters surface ethical tensions that remain less visible in textual or visual design fiction \cite{dunne2024speculative, cameraCars2023, CHI22VR, CHI24AIparent}. Within HFI, our approach builds on edible speculative interfaces and multisensory provocations \cite{grime08, CHI24FoodLeaf, CHI24foodFupop, deng2023dancing, deng2024}, suggesting that tangible encounters can serve as provisional “test sites” where hybrid food practices are rehearsed before they exist at scale.

Crucially, these encounters operated as ethical sandboxing rather than controlled evaluation. Participants developed and revised interpretive frameworks by drawing on market logics, ritual norms, AI ethics, posthumanist reasoning, and cultural taboos to model possible food futures \cite{jewitt_barker_steimle_2022, lupton_willis_2021}. In this mode, they were not merely reacting as informants; they were experimenting with ethical positions, imagining regulatory regimes, and anticipating cultural adaptations. Prior work on speculative participation and anticipatory governance suggests that publics can actively surface normative assumptions and propose alternative futures \cite{participatefood, ringfort2023, soctechimag, Forlano02102014}. Our findings resonate with this work: the micro-negotiations around responsibility (“who is accountable if a semi-living food system malfunctions?”), ritual (“how would hybrid foods be blessed, shared, refused?”), and agency (“does a semi-autonomous edible robot have a right not to be consumed?”) show publics performing ethical reasoning normally reserved for expert domains.

Taken together, speculative dining enacted a form of collective, embodied sandboxing in which participants tested moral intuitions, rehearsed social norms, and learned through sensory and relational trial-and-error. Rather than producing consensus, these interactions generated provisional, situational insights grounded in lived experience. This suggests a methodological implication for HCI, HRI, and HFI: when publics are engaged through sensory, material, and shared encounters, they co-produce anticipatory knowledge about emerging food technologies in ways that are exploratory, reversible, and ethically consequential. In this sense, the sandbox is not metaphorical but material---structured through the textures, movements, and rituals of speculative eating---and it enables publics to co-theorize futures in ways that are iterative, uncertain, and open-ended.

\section{\zm{Limitations and Future Work}}

 Our participant group was small and intentionally composed of individuals from creative and cultural sectors. This supported depth of reflection but limits the transferability of the findings. To address this, future work should engage broader publics---including culinary practitioners, technologists, ethicists, and general consumers---to examine how diverse communities interpret speculative biohybrid food systems and how speculative encounters vary across social worlds.

The scenario drew heavily on the culturally specific symbolism of the Ortolan ritual within a fictional 2052 Parisian setting. This framing likely influenced the meanings participants attributed to ritual, taboo, and hybrid organisms. Future studies could adapt the dinner-in-the-drama to different cultural contexts, exploring how food traditions, religious norms, and local symbolic systems shape public engagement with synthetic and hybrid forms of eating.

Although our findings highlight how multisensory tangible design fiction can support ethical and cultural reasoning, the study did not compare this method against alternative speculative modalities. To determine what is unique about embodied, sensory engagement, future research should systematically compare dinner-in-the-drama experiences with text-based scenarios, video design fictions, VR environments, or non-sensory roleplay to assess how different modalities elicit different forms of reflection.

Our speculative artifacts were supported by Wizard-of-Oz control and culinary proxies rather than functioning biohybrid systems. These choices ensured feasibility and participant safety but may have shaped perceptions of agency, materiality, and realism. Future work could explore higher-fidelity or interactive prototypes---such as soft robotics, edible robotics, bioinspired robotics (e.g., a flapping-wing drone \cite{inaflap}), biohybrid robotics, or tissue-mimicking constructs---to investigate how increasing realism affects ethical reasoning and sensory interpretation.
\section{\zm{Conclusion}}

\zm{This paper presented a tangible design fiction—a multisensory dinner-in-the-drama—that invited participants to confront the imagined consumption of edible biohybrid flying robots. By shifting speculation from narrative description to embodied encounter, the work revealed how technological, cultural, and ethical reasoning becomes intertwined when futures are experienced through taste, touch, sound, and ritual rather than imagined at a distance. Participants used the sensory and symbolic affordances of the scenario to speculate on technological pathways, reinterpret culinary traditions, and negotiate moral boundaries around hybrid organisms, surfacing perspectives often overlooked in conventional foresight. This work expands HCI’s repertoire for engaging publics with emergent biotechnologies, especially those poised to reshape intimate, everyday practices such as eating. As edible robotics and biohybrid systems might move closer to feasibility, the challenge for HCI is not only to evaluate their functionality but to understand how they will transform cultural meaning, social relations, and ethical worlds.}

\begin{acks}
We thank ArtX Gallery for providing the venue for the dinner theater event, and Ting Zhong for both photography and preparing the food served during the event. We are also grateful to all participants who voluntarily took part in this study. We acknowledge the Wallenberg AI, Autonomous Systems and Software Program -- Humanities and Society (WASP--HS). ZW’s and MF’s contributions were supported by the Marianne and Marcus Wallenberg Foundation, and NB’s contribution was supported by a Leverhulme Trust Early Career Fellowship (ECF-2021-065).
\end{acks}

\newpage
\bibliographystyle{ACM-Reference-Format}
\bibliography{base}

\newpage
\appendix

\begin{center}
    \huge \textbf{Appendices}\\
\end{center}

\section{\zm{Post-Event Interview Questions}} \label{sec:interview}

The following semi-structured questions were used to capture participants’ reflections after the dinner event. Interviewers selected relevant prompts flexibly depending on participants’ responses.

\subsection*{Overall Experience of the Event}
\begin{itemize}
  \item How would you describe your overall experience of participating in today’s dinner-in-the-drama?
  \item Did any moment feel particularly memorable, surprising, or meaningful?
\end{itemize} 

\subsection*{Sensory and Embodied Engagement}
\begin{itemize}
  \item How did you experience the food—its taste, appearance, smell, and presentation?
  \item How did the dramatic elements influence your immersion in the staged future?
  \item During the main course, what thoughts or feelings did you have while eating under the napkin?
\end{itemize} 

\subsection*{Interpretation of the Biohybrid Organism}
\begin{itemize}
  \item What were your impressions of the biohybrid prototype’s appearance?
  \item How did you experience the drone while it was flying? Could you imagine it functioning as a food-producing organism?
  \item How would you describe the identity of the Animal–Robot Hybrid (the “Robotic Ortolan”)? Did it feel more like an animal, a robot, or something else?
\end{itemize} 

\subsection*{Ethical, Cultural, and Sustainability Considerations}
\begin{itemize}
  \item What ethical considerations came to mind regarding: \\
  \{the biohybrid organism presented at the event, futuristic food systems more broadly, future bioengineering practices, future AI and robotics\}?
  \item Thinking about sustainability, to what extent do you feel the fictional technology was: \\
  \{environmentally sustainable, economically sustainable, socially sustainable\}?\\
  Why or why not in each case?
\end{itemize} 

\subsection*{Societal Adoption and Future Imaginaries}
\begin{itemize}
  \item What challenges or opportunities do you foresee in the broader adoption of:\\
\{the fictional biohybrid food system, future food technologies more broadly, future bioengineering, future AI and robotics\}?
  \item How soon do you think a future like the one depicted might emerge? What informs your view?
  \item Did the scenario feel more utopian, dystopian, or something in between? Why?
  \item Before the event, we provided an initial description of the dinner. How did your actual experience compare with your expectations?
\end{itemize} 

\begin{figure*}[h]
    \centering
    \includegraphics[width=\linewidth]{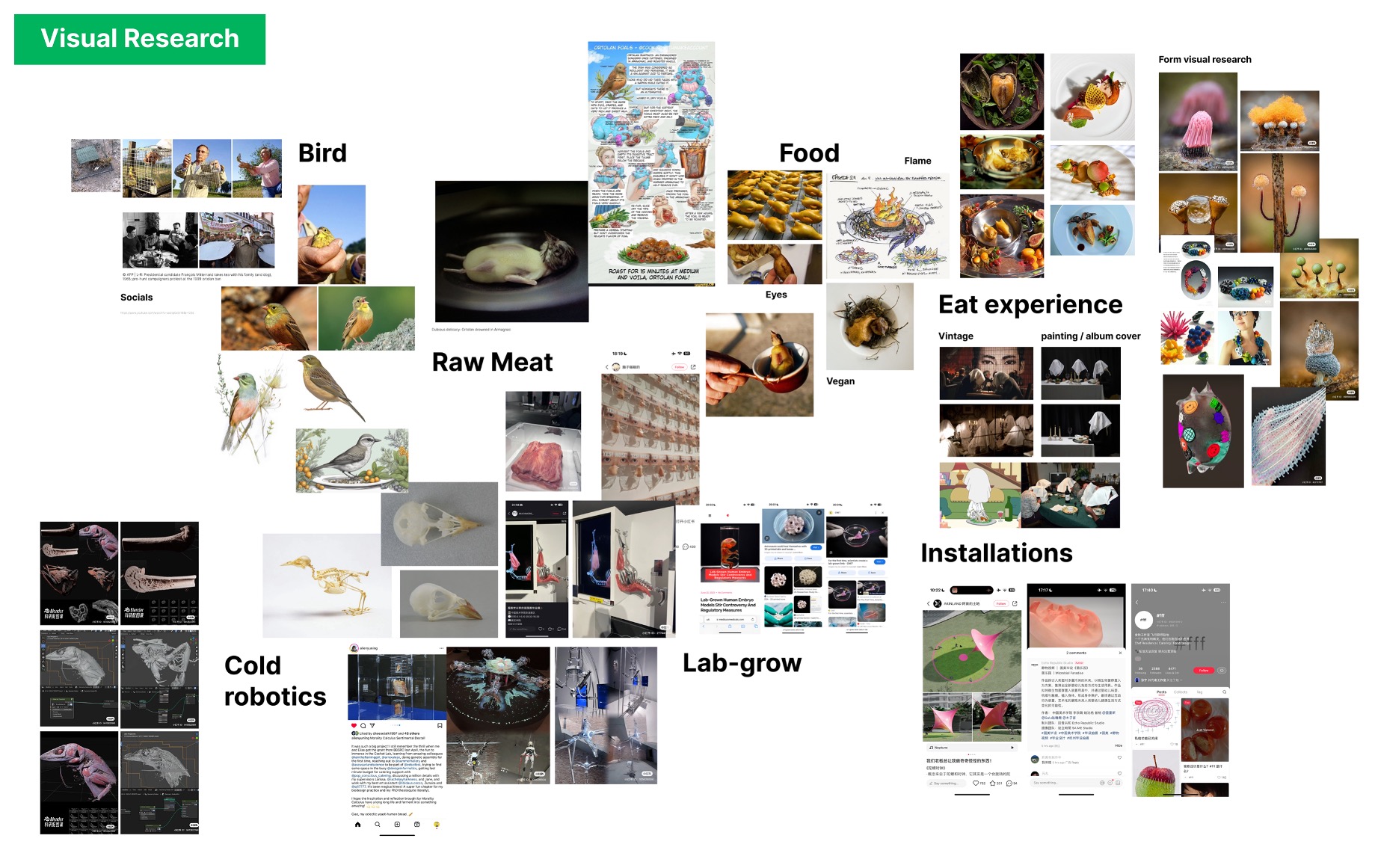}
    \caption{Vision research mapping conceptual dimensions of food-experience design relevant to the fictional biohybrid creature.}
    \Description{Visual research board collage organized into themes—Bird, Raw Meat, Food, Eat Experience, Cold Robotics, Lab-grown, and Installations—combining photographs, illustrations, lab imagery, food plating, sculptures, and social media screenshots to explore intersections of biology, consumption, technology, and aesthetics.}
    \label{fig:vision_appendix}
\end{figure*}

\section{Visual and Material Development}\label{sec:design_appendix}

This appendix documents the visual and material explorations that informed the development of the fictional biohybrid flying robot. These studies were instrumental not only in defining the creature’s aesthetic form but also in advancing our conceptual inquiry into how hybrid life might be interpreted through sensory and symbolic cues. Following research-through-design traditions, the sketches, models, and material tests presented here functioned as reflective probes through which we examined how the organism’s food-ness, machine-ness, and animal-ness could be made legible—even as these categories deliberately blurred.

\begin{figure*}[h]
    \centering
    \includegraphics[width=0.75\linewidth]{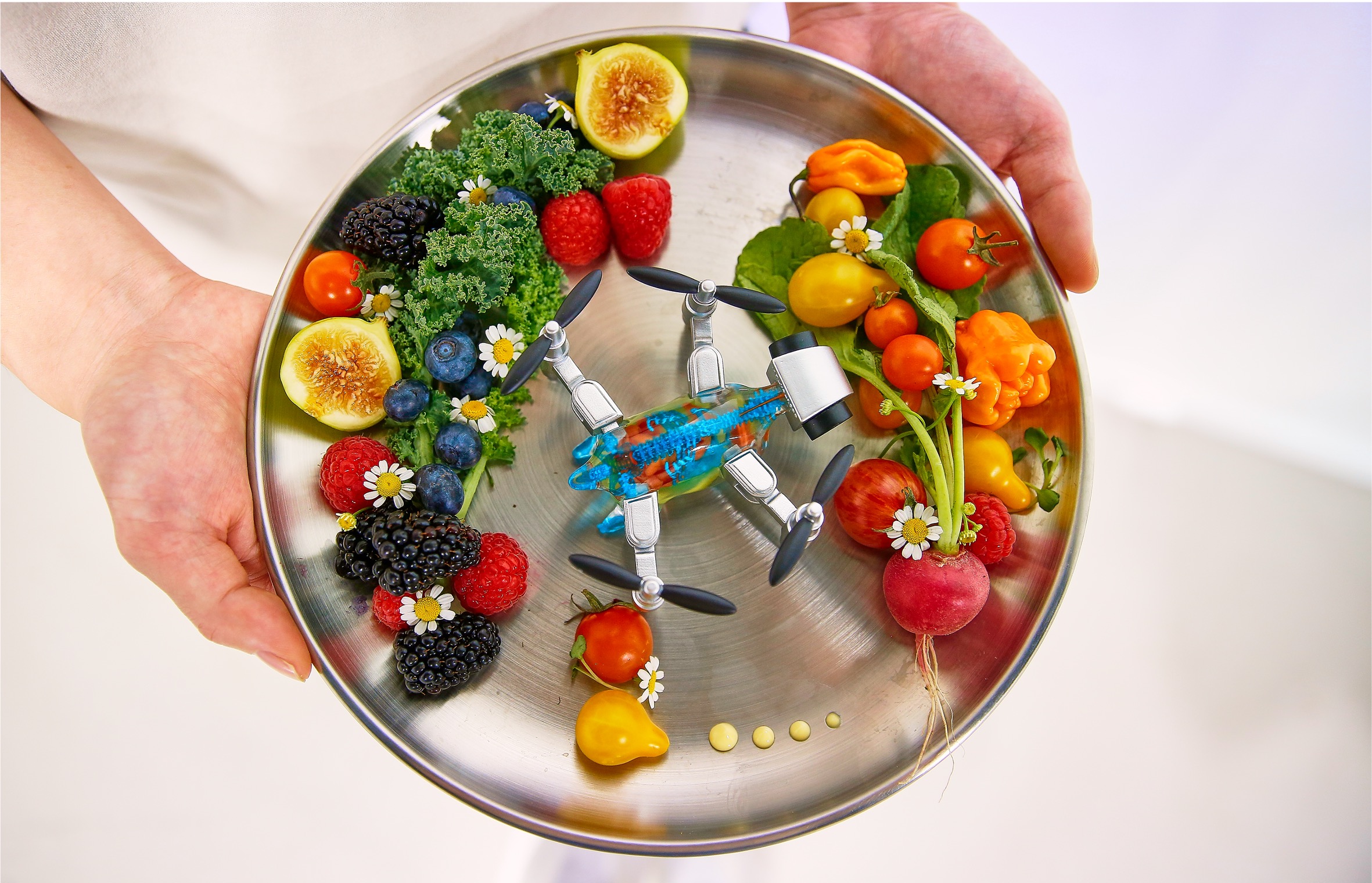}
    \caption{3D-printed prototype presented on a plate.}
    \Description{A pair of hands holds a round metal plate arranged with colorful fruits and vegetables in a circular pattern, with a small bird-shaped quadcopter model placed at the center like a plated dish.}
    \label{fig:bdop}
\end{figure*}

~\autoref{fig:vision_appendix} summarizes a preliminary mapping of themes from speculative gastronomy, edible robotics, and avian iconography. These mappings informed subsequent physical explorations of wings, head structures, sensory elements, and textural components, with the intent of outlining how the organism’s techno-organic identity might be conveyed visually.

\begin{figure*}[h]
    \centering
    \includegraphics[width=\linewidth]{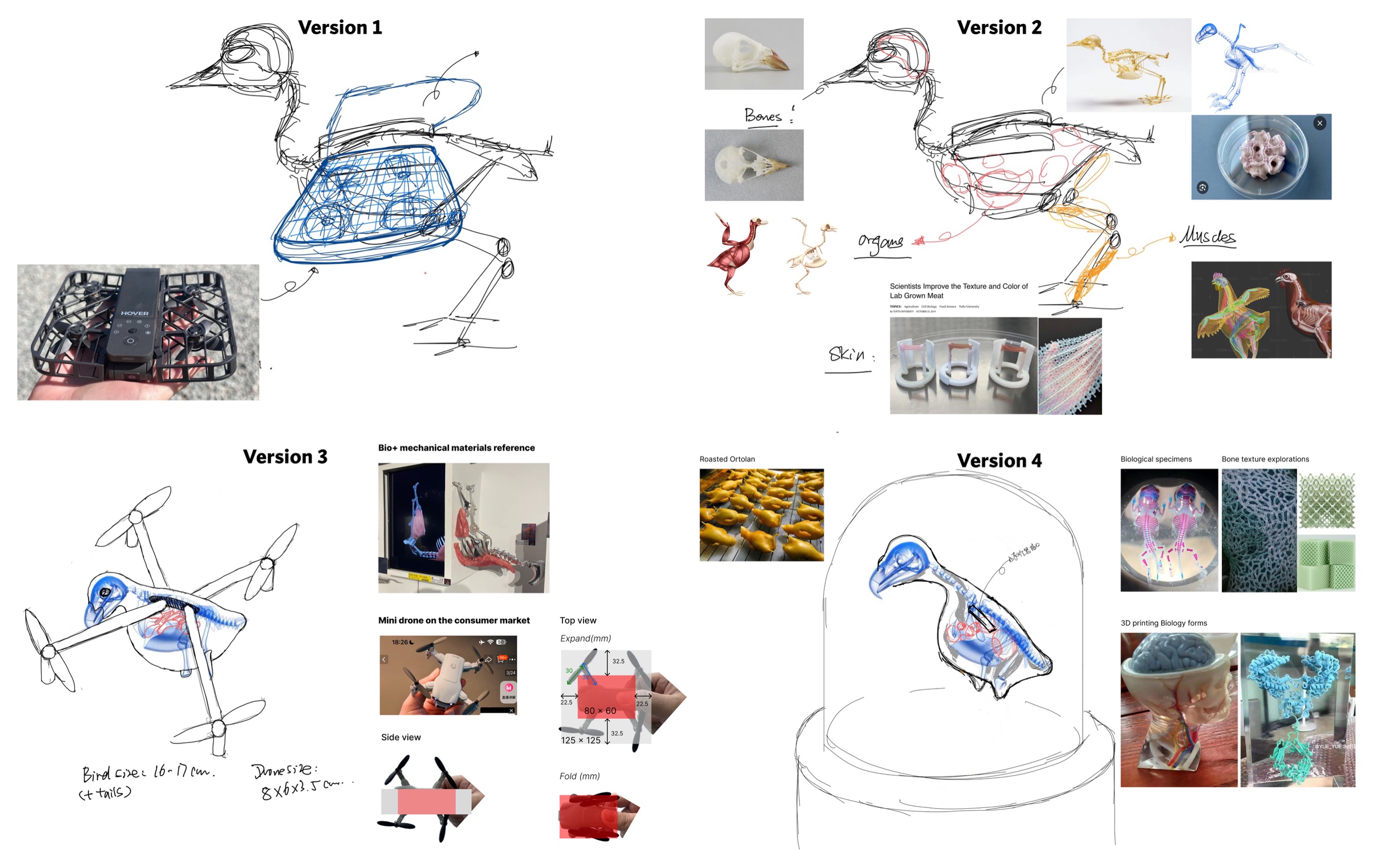}
    \caption{Wing iterations exploring transitions between avian morphology and hybrid propulsion geometries.}
    \Description{Concept development sheet showing four design versions of a bird-like robotic object: hand sketches overlaid with bones, organs, muscles, and skin annotations, alongside reference images of bird skeletons, lab-grown tissues, drones, biological specimens, food textures, and fabrication methods, illustrating the evolution from mechanical drone form to bio-inspired, specimen-like presentation.}
    \label{fig:wings}
    \vspace*{15pt}
\end{figure*}

The wing studies in ~\autoref{fig:wings} explore combinations of avian structures and propeller-inspired geometries. These iterations investigated how the fictional organism might reconcile visual cues of biological flight with mechanically augmented movement, contributing to the broader internal logic of its hybrid identity.

\begin{figure*}[h]
    \centering
    \includegraphics[width=\linewidth]{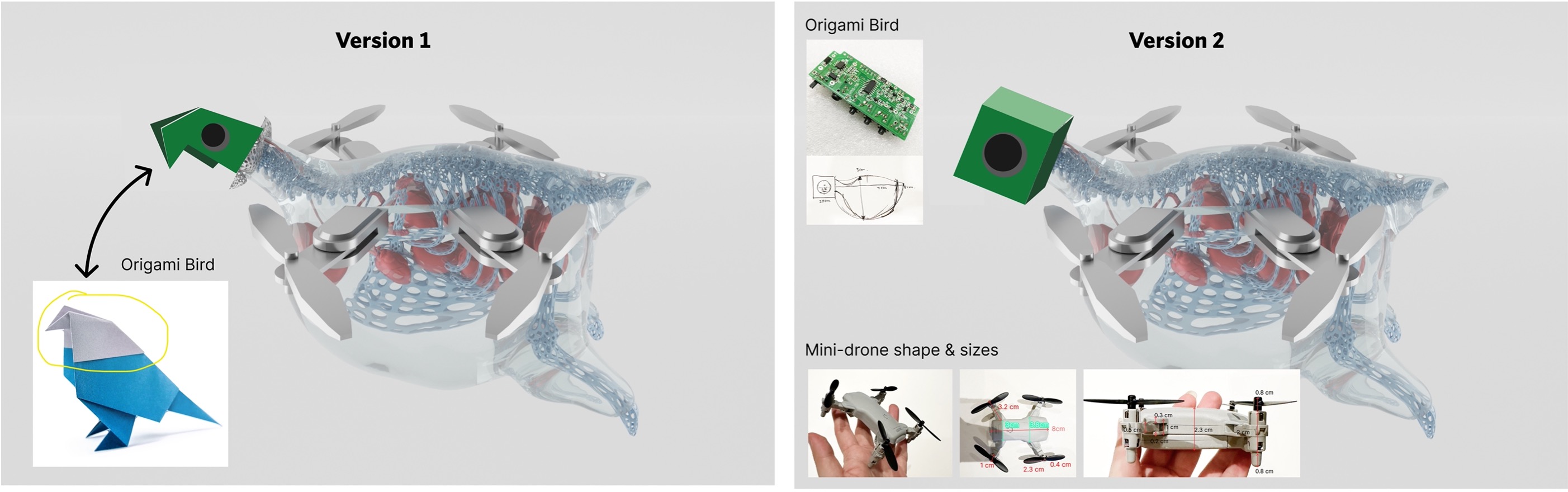}
    \caption{Head iterations examining sensory orientation and machinic expression.}
    \Description{Two versions of a bird-like robotic drone, showing a transparent body with internal bio-inspired structures and quad-rotor arms; Version 1 references an origami bird for head form, while Version 2 integrates a cube-shaped head and mini-drone electronics with size and shape references.}
    \label{fig:head_appendix}
\end{figure*}

The head experiments, as shown in \autoref{fig:head_appendix}, considered different ways of communicating non-biological perception and intelligence, including enlarged optical surfaces, flattened sensor planes, and stylized cranial forms. These variations informed decisions about how strongly to emphasize the artificial aspects of the hybrid within the final prototype.

\begin{figure*}[h]
    \centering
    \begin{subfigure}{0.77\linewidth}
        \centering
        \includegraphics[width=\linewidth]{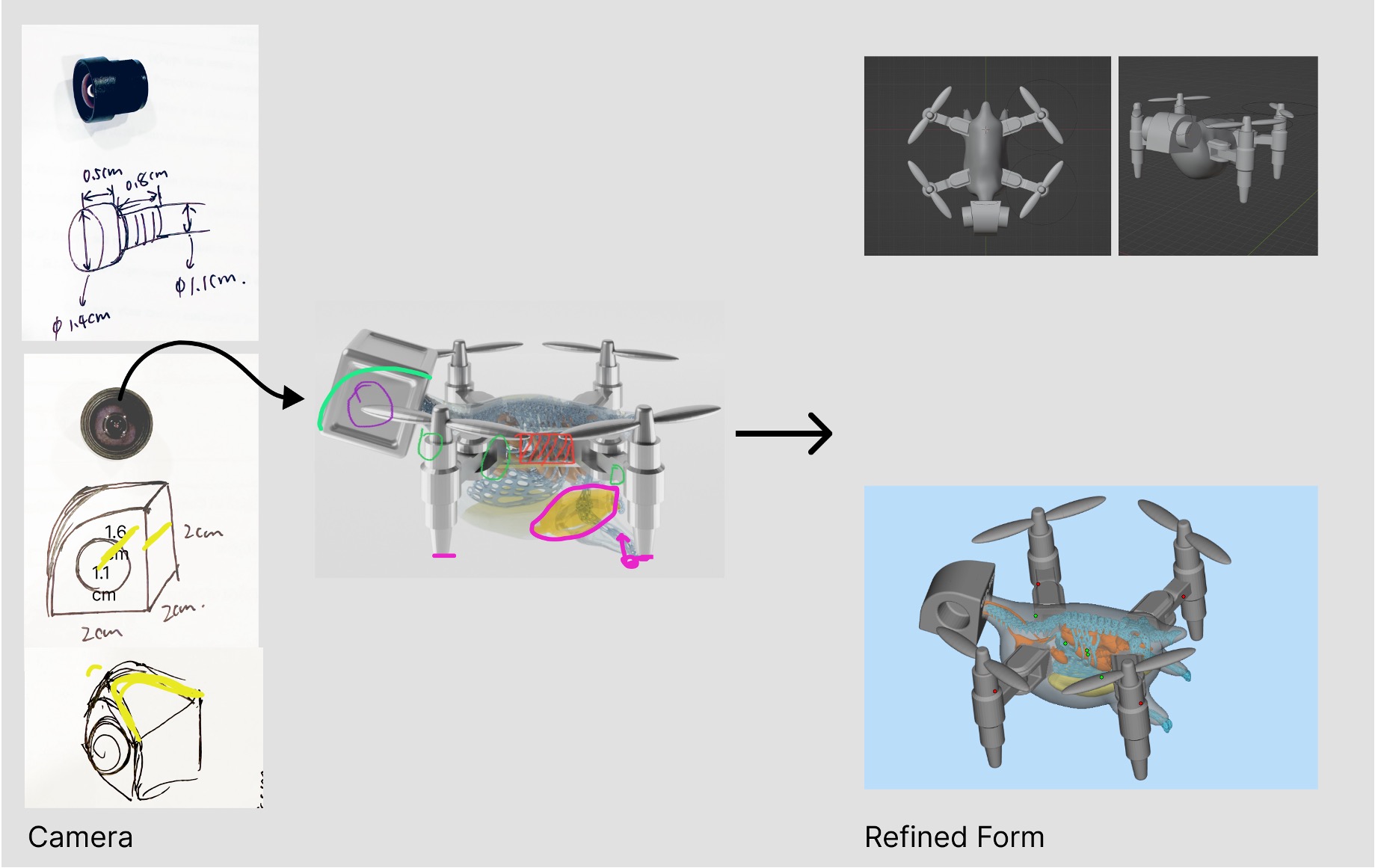}
        \caption{Eye and perception modules}
    \end{subfigure}\\
    \begin{subfigure}{\linewidth}
        \centering
        \includegraphics[width=0.96\linewidth]{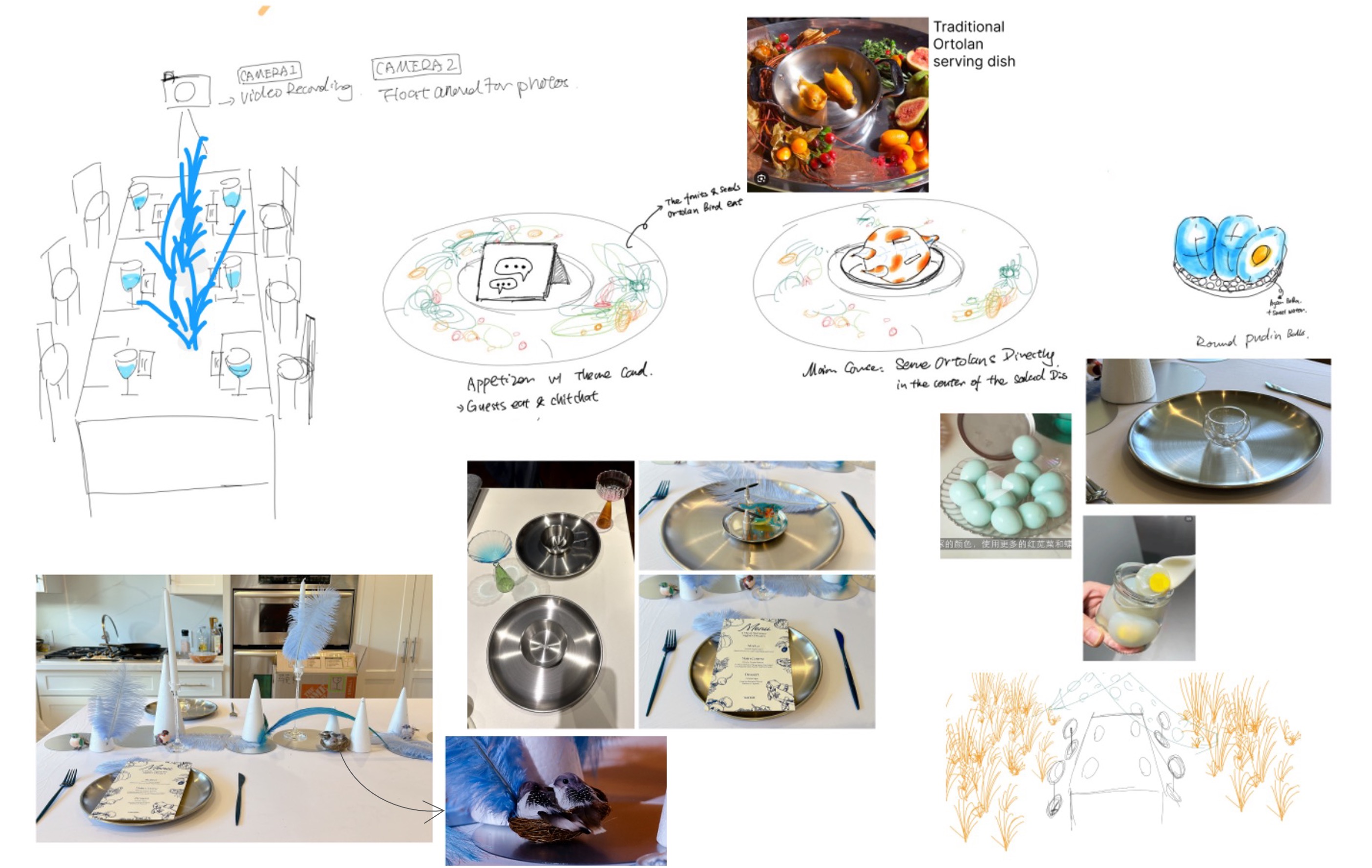}
        \caption{Feather–material, decorative birds, and nesting explorations}
    \end{subfigure}
    \caption{Additional form and texture explorations.}
    \Description{Two subfigures: (a). Perception and form-development diagram showing camera module references and dimension sketches feeding into a bird-like quadcopter design, with annotated component placement and a final refined 3D model integrating the camera into the bio-inspired robotic body. (b). Experience and table-setting concept collage combining hand sketches and reference photos: a multi-course dining scenario inspired by the traditional ortolan serving ritual, showing plate compositions, camera placement for documentation, sculptural table decor, plated dishes, and bird-themed visual motifs.}
    \label{fig:details_appendix}
\end{figure*}

Additional studies (\autoref{fig:details_appendix}) explored how perception modules, feather-material, decorative birds, and nest-related textural elements might signal techno-organic identity. They contributed to the conceptual consistency of the creature’s fictional ecology and informed decisions about its rendered material presence. \autoref{fig:bdop} presents the 3D-printed prototype theatrically staged on a plate to explore the performative presentation of the props.

\begin{figure*}[h]
    \centering
    \includegraphics[width=0.72\linewidth]{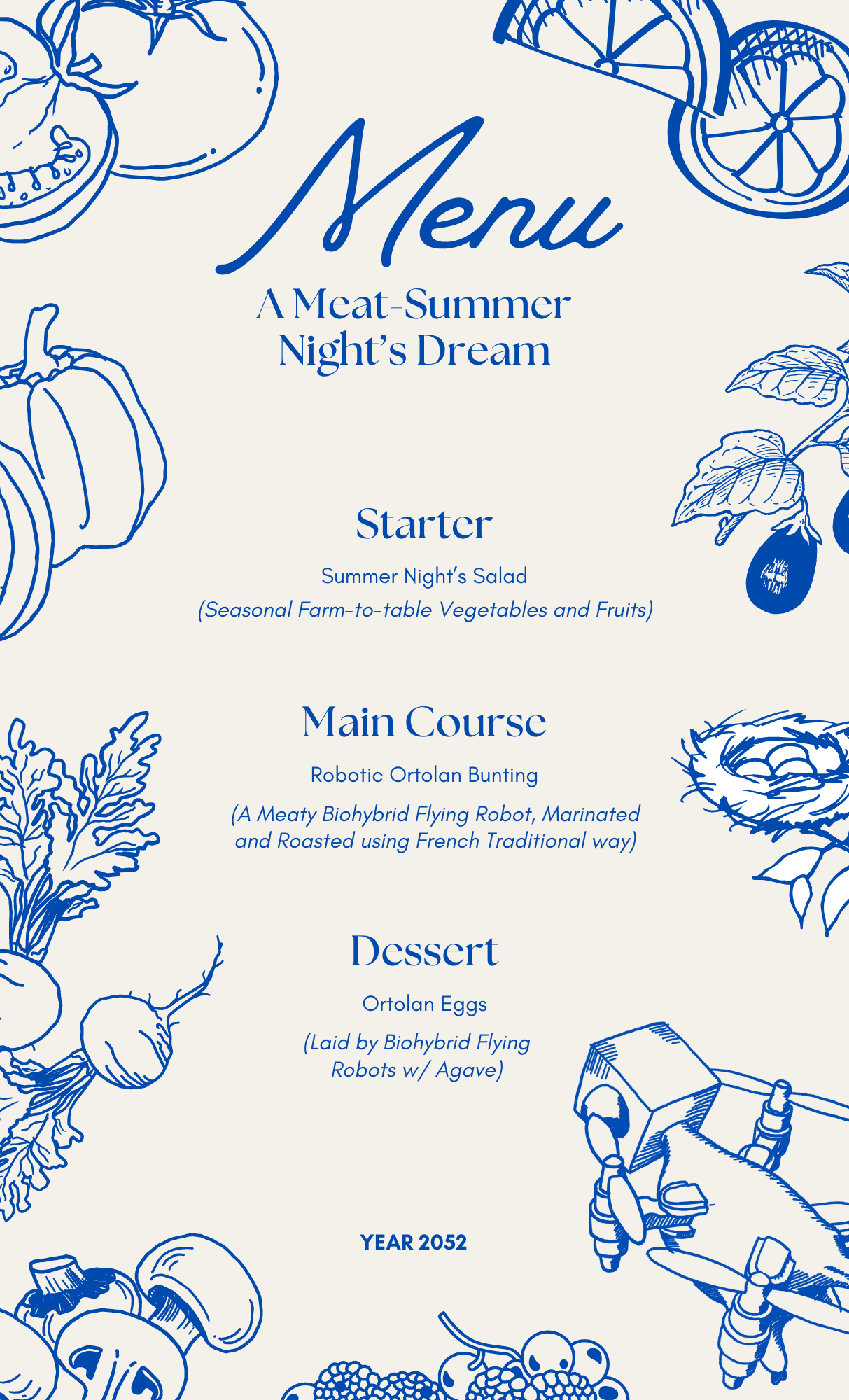} 
    \Description{A menu with illustrations printed in blue.}
    \caption{The menu.} 
    \label{fig:menu}
\end{figure*}

\autoref{fig:menu} shows the graphic design of the menu, including textual information about the three-course dinner and sketches of elements from the dinner theater.

\section*{}
\vfill
\begin{displayquote}
\textit{``Are you sure\\ That we are awake? It seems to me, \\That yet we sleep, we dream. ''}    \\ \hspace*{\fill}— William Shakespeare,\\ \hspace*{\fill} A Midsummer Night’s Dream. \\

\end{displayquote}

\end{document}